\documentclass[A4paper,12pt]{article}
\usepackage[margin=2.4cm,nohead]{geometry}
\usepackage{graphicx}
\usepackage{amsfonts}
\usepackage{amssymb}
\usepackage{amsmath}
\usepackage[english]{babel}
\usepackage{appendix}
\usepackage{lscape}
\usepackage{subcaption}
\usepackage{caption}
\usepackage{hyperref}
\usepackage{natbib}
\usepackage{setspace}
\usepackage{color} 
\usepackage{tabularx}

\newcommand{\rl}{\noindent\hangindent=0.5cm}
\newcommand{\ed}{\end{document}}
 
\begin{document} 

\title{Sorting on the Used-Car Market After the Volkswagen Emission Scandal\thanks{We benefitted from comment by Stefan Bühler, Nicolas Eschenbaum, and Philipp Zahn. We are indebted to Andreas Hermann and Benjamin Polak for their data provision and support. The usual disclaimer applies.}}

\author{Anthony Strittmatter\footnote{SEW, University of St.Gallen; email: anthony.strittmatter@unisg.ch.} \and Michael Lechner\footnote{SEW, University of St.Gallen; CEPR, London; CESIfo, Munich; IAB, Nuremberg; IZA, Bonn.; email: michael.lechner@unisg.ch.}}

\date{\today}

\maketitle
\doublespacing

\begin{abstract}
\noindent 
The disclosure of the VW emission manipulation scandal caused a quasi-experimental market shock to the observable environmental quality of VW diesel vehicles. To investigate the market reaction to this shock, we collect data from a used-car online advertisement platform. We find that the supply of used VW diesel vehicles increases after the VW emission scandal. The positive supply side effects increase with the probability of manipulation. Furthermore, we find negative impacts on the asking prices of used cars subject to a high probability of manipulation. We rationalize these findings with a model for sorting by the environmental quality of used cars. \medskip

 \medskip

\end{abstract}

\newpage
\section{Introduction}\label{sec1}

Automobile emissions are partly responsible for air pollution and global warming. If car owners care about the emissions of their cars, we expect the used-car market to react to the arrival of new information about the emissions of cars. One possible reaction is sorting based on preferences for environmental quality \citep[see, e.g.,][for a comprehensive discussion of sorting in markets for durable goods under heterogeneous preferences]{he99}. Car owners with high preferences for environmental quality could sell their cars to individuals with low preferences for environmental quality.

In this study, we exploit the decline in the observed environmental quality of Volkswagen (VW) diesel cars after the disclosure of the stunning VW emissions scandal to investigate reactions to information disclosure. On September 18, 2015, the US Environmental Protection Agency (EPA) disclosed the installation of defeat device software in the emission control systems of VW diesel engines \citep[][]{ep15}. This disclosure reduced the observable environmental quality of VW diesel cars because the defeat device negatively affects the environmental and engine performance of the manipulated cars. Furthermore, the durability of several parts of the exhaust system (e.g., the catalytic converter) could significantly decline after retrofitting, thus increasing maintenance costs. Since market participants did not expect the information revealed by this disclosure, the resulting decline in observable environmental quality is a quasi-experimental exogenous shock to the used-car market.

Several studies show that car buyers' willingness to pay is sensitive to fuel prices \citep[e.g.,][]{al13,bu13,sa16}. These studies argue that car buyers take the future fuel costs of vehicles with different fuel economies into account. In contrast to these studies, we investigate the reactions of potential car sellers to information disclosure about the environmental quality of cars, which is related to fuel economy.

\cite{ta15} investigate the random disclosure of information (for potential bidders) about the quality of used cars during online auctions. The unexpected disclosure of new information increases sales probability and revenue for sellers. Surprisingly, the results remain the same regardless of whether the disclosed information increases or decreases the observable quality. They argue that the (re-) sorting of bidders with heterogeneous quality preferences increases competition, even if the new information reveals low quality. In contrast to \cite{ta15}, we investigate a sorting mechanism on the supply side and focus on aspects related to environmental quality.

\cite{pe14} investigate the defect probability of different car components and the corresponding turnover rates for different car models. Components with a high probability of an easily detectable defect increase the turnover rates of cars. This is evidence for sorting on the supply side with respect to observable car quality. However, it is difficult to test the causal effect of information disclosure on the supply of used cars because we are usually unable to observe exogenous variation in observable used-car quality. Furthermore, \cite{pe14} do not consider car components that affect the environmental quality of cars.

We base our analyses on data collected from a large German online car advertisement portal. The volume of this online market is approximately 3.6 million cars per year, which reflects a trade value of 40 billion euros per year or 50\% of the German used-car market.\footnote{\rl The German Federal Motor Transport Authority (Kraftfahrtbundesamt) documents 3 million first-time registered cars and 7 million ownership transfers of cars during the year 2014 \citep{kr17}.} Our sample contains 1.1 million newly offered cars between August 2015 and April 2016. Furthermore, we observe detailed car and seller characteristics. We employ difference-in-differences estimators to investigate asking prices and market shares of diesel cars by carmakers and time periods before and after the disclosure of the VW emission scandal. In particular, we use unconditional and conditional differences-in-differences estimators. In the latter approach, we account for important variables that could affect the structure of the used-car market. For this purpose, we use semi-parametric radius matching estimators with bias adjustment \citep{le11}.

We find that the supply of used VW diesel cars increases significantly after the disclosure of the emission manipulation scandal. The positive supply effects increase with the probability of manipulation. Furthermore, we find evidence of negative effects on asking prices for cars with a high probability of manipulation. We rationalize these findings using Peterson and Schneider’s (2016)\nocite{pe16} model of sorting in durable goods markets.

We provide evidence that spillover effects to non-manipulated VW diesel cars do not explain an increase in the supply or decline in the asking price of used cars. This mitigates the plausibility of several alternative explanations for our findings, for example, that the overall reliability of diesel cars decreases after the disclosure of the scandal or that mainly risk averse car owners bring their used cars to market.

Furthermore, we provide evidence that these results are strongly driven by the supply of minivans and sport utility vehicles (SUVs). Both vehicle classes are heavy polluters. Compact cars drive the negative asking price effects. Finally, we show that it is mainly professional car dealers who do not offer a warranty that increase their supply of VW diesel vehicles after the disclosure of the VW emission scandal.

In the next section, we introduce the background of the VW emission scandal. In Section \ref{sec3}, we introduce a small sorting model. In Section \ref{sec4}, we describe the data. In Section \ref{sec5}, we describe our empirical strategy. In Section \ref{sec6}, we report the empirical results. In Section \ref{sec7}, we discuss alternative explanations for our findings. The final section concludes the paper. We provide additional material in Online Appendices A-E.

\section{The VW emission scandal \label{sec2}}

Many countries use emission standards to limit toxic and greenhouse gases in automobile exhaust. In the last two decades, the European Union and the US significantly reduced the nitrogen oxide (NOx) limits for automobile exhaust (see Online Appendix A for details). Car manufacturers had to improve the emission control technology in their cars. In 2007, the US authorities certified the VW diesel engine EA 189 under the strictest nitrogen oxide limits at the time. The engine achieved low nitrogen emissions with a nitrogen oxide storage catalytic converter \citep[see, e.g.,][for details]{ya15}. This converter stores nitrogen oxide in a catalyst material (often barium) and converts it into nitrogen (N2) and carbon dioxide (CO2). The conversion process requires the injection of additional fuel to regenerate the catalytic converter, which increases fuel consumption. The converter has a low durability, especially when the injected fuel includes high amounts of sulfur.\footnote{\rl An alternative technology is the selective catalytic reduction converter, which injects urea to lower nitrogen oxide emissions. Urea is sold under the brand name Ad Blue. Ad Blue has to be refilled regularly.}

However, the EA 189 engine has a defeat device for the emission control system. The defeat device software recognizes when a car is on the test stand. In test stand mode, the emission control system operates optimally. However, the capacity of the emission system is reduced or completely switched off when the car is not on the test stand \citep[see, e.g.,][]{th14}. VW sold 11 million cars worldwide with defeat devices (2.4 million cars in Germany).

Before the EPA disclosed the installation of the defeat device software, only a small number of managers and EPA officers were aware of the emission test manipulations \citep{cl17}. The nitrogen oxide emissions of cars with a defeat device exceed official limits by up to 40 times \citep{ne15}. Within days of the disclosure, VW shares lost more than 20\% of their value (approximately 15 billion euros), and the CEO of VW (Martin Winterkorn) resigned \citep{st17}. Since October 2, 2015, VW customers have been able to check an online portal to see if their car has a defeat device. On October 30, 2015, VW submitted retrofitting plans for cars in Germany, but considerable uncertainty remains. Although retrofitting is mandatory in Germany, many cars still have the defeat device. It appears that some cars only require a software update, while other cars need a serious engine retrofit.

On April 23, 2016, \textit{Der Spiegel} headlines revealed that Bosch AG (a German automotive supplier) delivered the defeat devices not only to VW but also to many other car producers worldwide \citep{sp16}.\footnote{\rl \textit{Der Spiegel} is an important weekly news magazine in Germany.} On October 25, 2016, US courts and VW agreed on a compensation payment of 15 billion US dollars to US customers \citep[approximately 480,000 cars with defeat devices were sold in the US; see, e.g.][]{st17}. This includes the costs of retrofitting, buybacks, and individual compensation of between 5,000 and 10,000 US dollars.

To date, there is still no clear decision about the obligation of VW towards misled customers in Germany. German laws do not force VW to repurchase diesel cars or to pay notable amounts of individual compensation. Nevertheless, VW has negotiated with various government institutions about retrofitting plans and other remedies. The latter usually consist of free software updates for cars with defeat devices. In exceptional cases, hardware components are also retrofitted. VW covers the costs of retrofitting but not necessarily of the increased maintenance costs due to retrofitting. German courts have had to address mass complaints by VW customers, but there was no clear court decision as of mid-2019.
The disclosure of the VW emission scandal affects the observable quality of VW diesel cars. The defeat devices could directly affect the environmental performance, power, and/or fuel consumption of the vehicle. Owners of manipulated cars must visit the garage for retrofitting, which is time consuming. Moreover, the durability of the exhaust gas recirculation valve, the particulate filter, and the catalytic converter could significantly decline after retrofitting. It appears that VW car owners have to bear at least some of the (direct or indirect) costs.

The image of VW diesel cars declined after the disclosure of the emission manipulation scandal \citep{cl17,st17}. Furthermore, the emission scandal could have had spillover effects on VW gasoline cars, as well as diesel cars from other producers \citep{cl17}. For example, the image of diesel cars could have declined, and individuals could have updated their expectations about future car quality fraud revelations. We assume that the decline in observable quality is stronger for VW diesel than for other carmakers, such that the relative loss in observable quality is negative for VW diesel cars.

Table \ref{tab1} shows trends in the German car market between 2008 and 2017. Since 2013, registrations of new gasoline and diesel cars have increased. This trend continued for gasoline cars after the disclosure of the VW emission scandal, but the registrations of new diesel cars remained stable in 2015 and 2016 and declined in 2017. The number of gasoline car ownership transfers has been stable since 2008. In contrast, the number of diesel car ownership transfers exhibits an increasing trend, which was not interrupted after the disclosure of the VW emission scandal. Owners decommission their cars because they want to either scrap or sell them at a later point in time. While the share of gasoline car decommissions remains stable over time, we observe an increasing trend in the decommissioning of diesel cars that was not interrupted by the VW emission scandal. It appears that the VW emission scandal had a clear, negative impact on the new car market for diesels, but it is difficult to draw conclusions about the used car market from this macro perspective.

\[ \text{Table \ref{tab1} around here}  \]

\section{Sorting \label{sec3}}

We use a simplified version of Peterson and Schneider's (2016)\nocite{pe16} model for sorting in durable goods markets to obtain a rationale for the potential effects of the VW emission scandal on the used car market \citep[see also][]{he99,he02}. The parameters of the model are as follows: Potential used-car sellers own one used car each. Potential used-car buyers are not endowed with any car. A potential used-car seller has the taste for quality $\theta$ that is drawn from the uniform distribution $[1,\theta_H]$. The cumulative distribution of potential used-car sellers with taste for quality $\theta$ is $H(\theta)=(\theta-1)/(\theta_H-1) $. All potential used-car buyers have the taste for quality $\theta=1$. The taste for the quality of potential buyers is lower than that of potential sellers, which reflects their willingness to buy cars that have been discarded by the potential car sellers. We assume that the number of potential used-car buyers exceeds the number of potential used-car sellers.

Let $Q>0$ be the quality of a used car (including environmental quality), which is observable for potential sellers and buyers. The utility of owning a car for a driver with taste $\theta$ is $\theta Q$. Drivers gain zero utility from owning a second car. Potential sellers have the option to replace their used car with a new car. The quality of a new car is $Q_N$ with $Q_N>Q$. The price of a new car is $P_N$. We assume that the new car market is competitive, such that $Q_N$ and $P_N$ are exogenous and not influenced by the used-car market.\footnote{\rl This is a strong simplifying assumption. However, if an interaction between the new and used car market biased our empirical results, then we would expect to find positive supply effects for manipulated and non-manipulated VW diesel vehicles. We show that the positive supply effects occur only for cars with a high manipulation probability, which mitigates concerns about such a bias.} Furthermore, we assume that $\theta_H Q_N>P_N>Q_N$, which implies that it is efficient for some potential sellers to replace their used car with a new one. Because we assume that the number of potential used-car buyers exceeds the number of potential used-car sellers, the market price of used cars equals the potential buyers' reservation prices, which are proportional to the observed car qualities. The used-car market price is $P(Q)=Q$ because $\theta=1$ for potential buyers of used cars. Accordingly, the marginal effects of observed quality on used-car prices are positive ($\partial P(Q)/ \partial Q=1$). This suggests that a lower observable environmental quality would have a negative impact on the prices of used cars.

Potential sellers have an incentive to sell their used car and buy a new car when the utility from driving the new car, $\theta Q_N$, minus the price differential between the prices of the new and the used cars, ($P_N-P(Q)$), exceeds the utility from driving the used car, $\theta Q$. The utility of trade is $u(Q)= \theta (Q_N-Q)-(P_N-Q)$. The marginal effect of the observed quality on the utility of trade is negative,
\begin{equation*}
\frac{\partial u(Q)}{\partial Q} = 1-\theta < 0.
\end{equation*}
This suggests that a decline in observed environmental quality would increase the utility of trade. This would imply that an owner of a VW diesel vehicle generates more utility from trade after the emission scandal. Potential sellers offer their car on the used car market when the utility of trade is positive. This suggests that a decline in the observable environmental quality increases the probability of supplying used cars.

\section{Data \label{sec4}}

We collect our data from a large German online car advertisement platform. Private car owners and firms offer their cars for sale on this platform. They usually post several pictures and detailed descriptions of the offered car together with an asking price. Potential buyers can search for cars on this platform. When they find an appropriate car, they contact the seller directly. The seller and potential buyer may arrange a meeting, which gives the potential buyer the possibility to inspect the car. They then negotiate mutually over the transaction price, which can deviate from the asking price.\footnote{\rl We do not observe transaction prices in the data.} If the seller and buyer agree on the terms of the trade, then they legally transfer ownership of the car.

\subsection{Selection of the estimation sample}

Our sample covers the daily inflow of used cars from August 26, 2015, to April 15, 2016. The sample period ends one week before Der Spiegel headlines revealed that Bosch AG delivered defeat devices to many car producers worldwide \citep{sp16}. We analyze only used cars with diesel or gasoline engines. We omit cars with a mileage above 200,000 km or that are older than 20 years. Table \ref{tab2} shows the five different vehicle classes we consider. The inflow sample contains 1,108,566 vehicles. The sample consists mainly of compact and medium-sized vehicles. We focus on nine common carmakers in Germany. The inflow sample includes 427,286 (39\%) VW, 182,393 (16\%) BMW, 136,613 (13\%) Opel, 136,393 (12\%) Mercedes-Benz, 134,200 (12\%) Ford, 33,585 (3\%) Toyota, 25,703 (2\%) Renault, 19,585 (2\%) Peugeot, and 12,808 (1\%) Fiat vehicles.

\[ \text{Table \ref{tab2} around here}  \]

\subsection{Descriptive Statistics}
Table \ref{tab3} shows the descriptive statistics for vehicles from VW and other carmakers. The share of diesel vehicles is slightly above 50\%. The first asking price does not differ between VW and other cars. The offered VW cars are somewhat newer than the offered vehicles from other carmakers. The average age of VW cars is three years, and the average mileage is approximately 60,000 km. The average age of the other carmakers is almost 4 years, and the average mileage is approximately 67,000 km; 76\% of the VW cars and 61\% of the vehicles from other carmakers have a full service history. Approximately 10\% of the sellers offer a warranty to purchasers. In Germany, new cars undergo a general inspection after three years and must revisit the general inspection every subsequent second year. Approximately 20\% of the offered cars have a new general inspection, which is valid for two years. For another 57\% of the VW cars and 49\% of the vehicles from other makers, the duration until the next inspection becomes due is more than one year. The share of cars offered by private sellers is approximately 10\%. Most car bodies are limousines or estate cars. The VW cars in the sample are more often regulated under the more recent emission standards than the vehicles from other carmakers, which reflects the younger age of the VW cars in the sample. Some German cities have environmental zones. Cars are only allowed to enter these zones when they have a special tag on the windshield that indicates low particulate matter emissions. The lowest particulate matter emissions is indicated by a green tag. Over 80\% of the offered cars have such a green particulate matter tag.

\[ \text{Table \ref{tab3} around here}  \]

\subsection{Time trends}
\begin{figure}[h!]
\caption{Inflow of used diesel vehicles by car make.} \label{fig1}
\centering
\fbox{\includegraphics[width=0.6\textwidth]{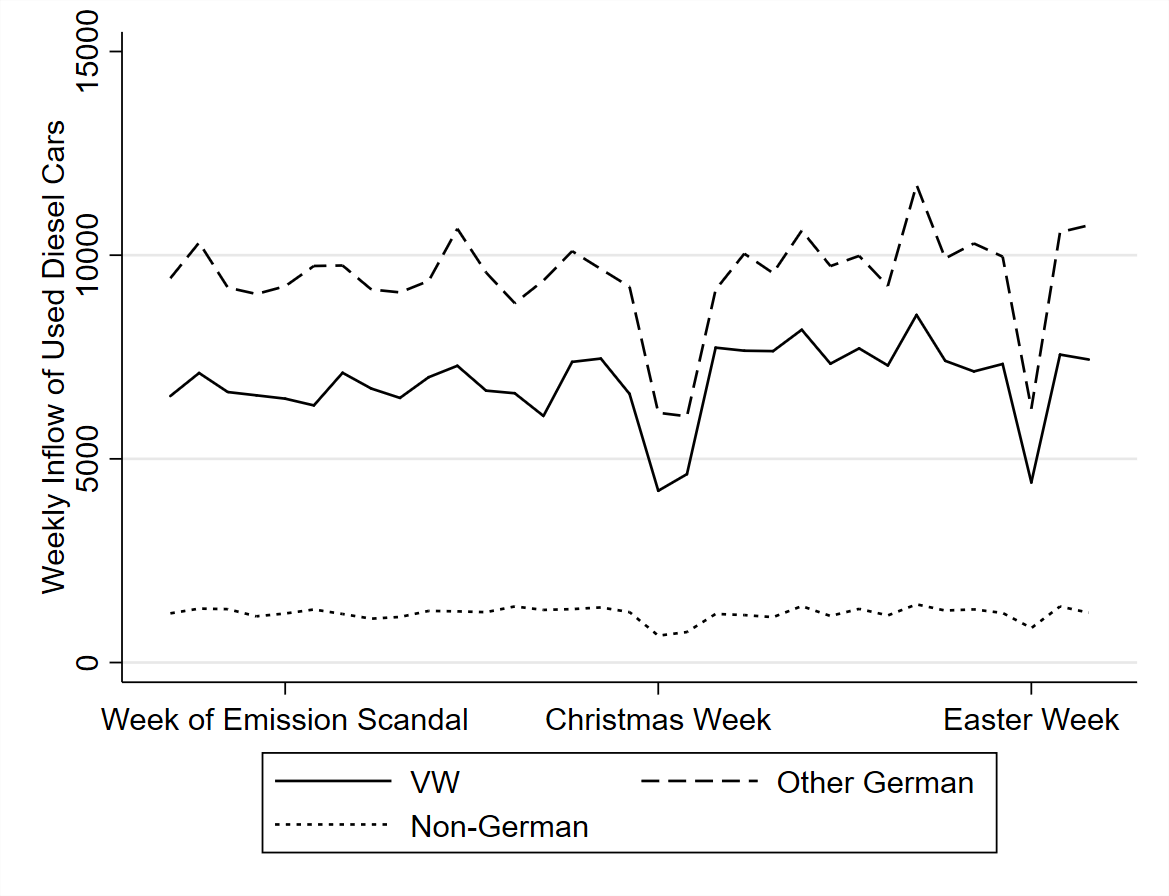} }
\parbox{\textwidth}{\footnotesize \emph{Notes:} This figure plots the weekly inflow of used diesel vehicles by car make. Other German car makes include Mercedes, BMW, Opel, and Ford. Non-German car makes include Renault, Peugeot, Fiat, and Toyota. }
\end{figure}

\begin{figure}[h!]
\caption{Inflow of used gasoline vehicles by car makes.} \label{fig2}
\centering
\fbox{\includegraphics[width=0.6\textwidth]{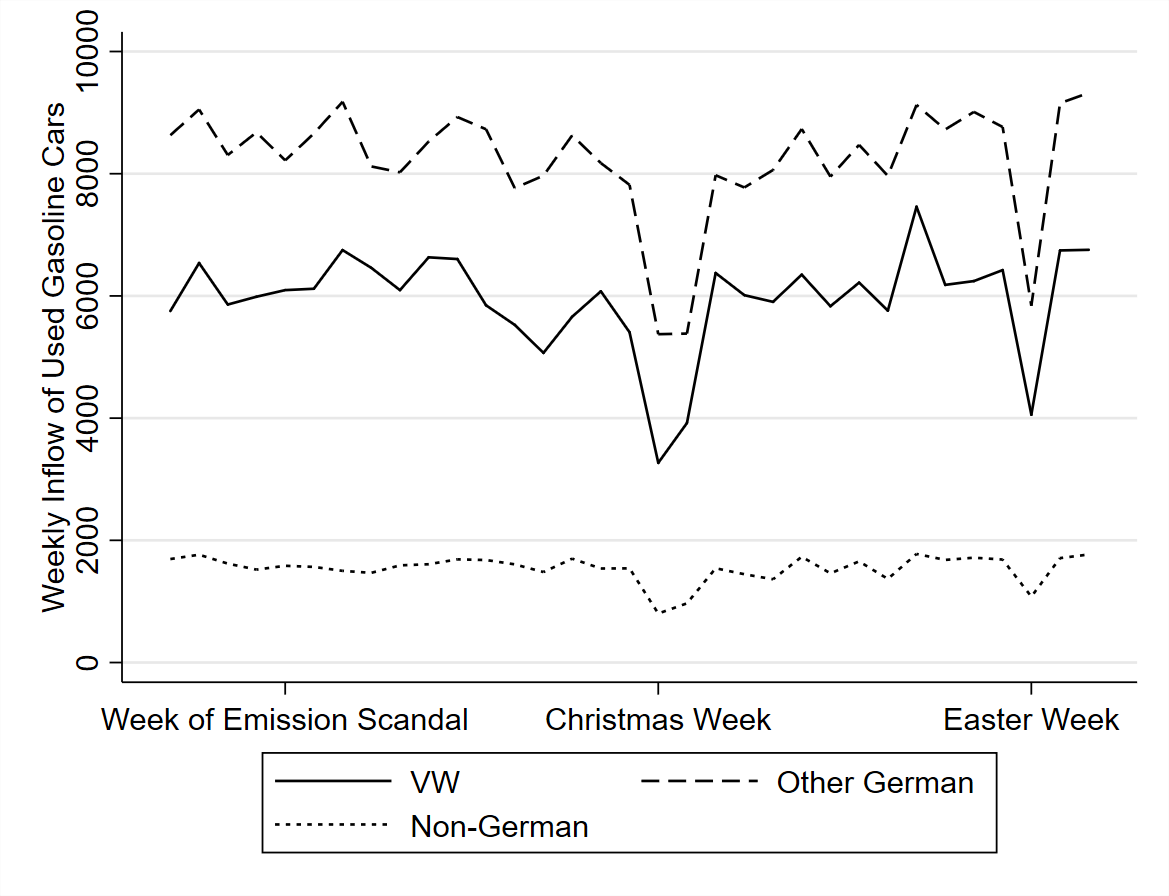} }
\parbox{\textwidth}{\footnotesize \emph{Notes:} This figure plots the weekly inflow of used gasoline vehicles by car make. Other German car makes include Mercedes, BMW, Opel, and Ford. Non-German car makes include Renault, Peugeot, Fiat, and Toyota. }
\end{figure}

\begin{figure}[h!]
\caption{Share of inflowing used diesel vehicles by car make.} \label{fig3}
\centering
\fbox{\includegraphics[width=0.6\textwidth]{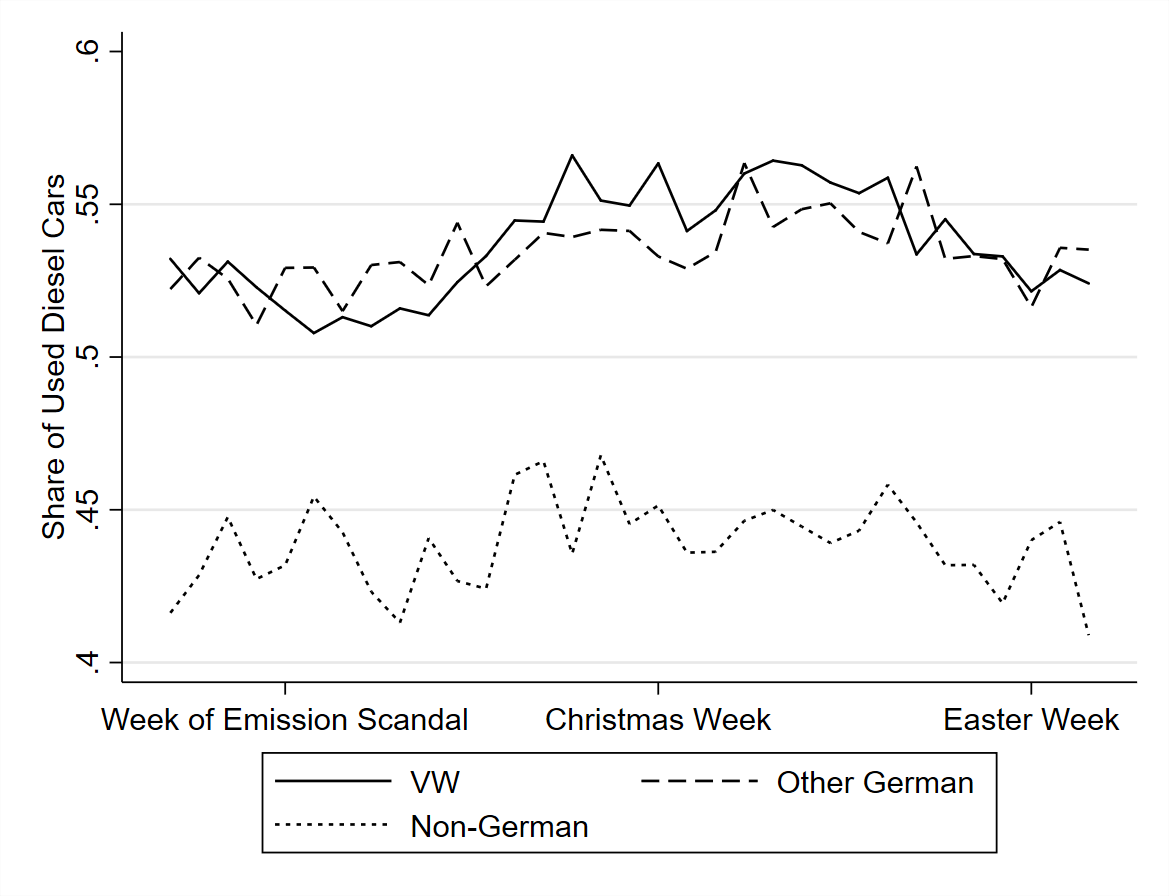} }
\parbox{\textwidth}{\footnotesize \emph{Notes:} This figure plots the weekly share of inflowing used diesel vehicles by car make. Other German car makes include Mercedes, BMW, Opel, and Ford. Non-German car makes include Renault, Peugeot, Fiat, and Toyota. }
\end{figure}

Figure \ref{fig1} shows the number of newly incoming diesel cars used per week. We distinguish among VW, other German, and non-German car makes. We observe high seasonal fluctuations, especially around Christmas and Easter. Figure \ref{fig2} reports a similar time pattern for the inflow of used gasoline cars. Accordingly, a comparison of the inflow of used cars before and after the emission scandal would capture these seasonal fluctuations. In contrast, Figure \ref{fig3} documents that the share of newly incoming diesel cars is more stable over time by car make. The share of non-German diesel cars appears to be stable over time. The share of VW diesel and the share of other German diesel cars have humped shaped patterns after the VW emission scandal. The humped shaped pattern is more pronounced for the share of VW than other German diesel cars.

\[ \text{Figures \ref{fig1}, \ref{fig2}, and \ref{fig3} around here}  \]

The average first asking price is 16,732 euros, and the average final asking price is 16,350 euros. Accordingly, the average discount is 381 euros or 2\% of the first asking price. The first and final asking prices differ for only 31\% of the offered cars. We focus on the first asking price because given the timing of our data, the final asking price could already be contaminated (when the car was online before the disclosure of the VW emission scandal and sold afterwards). Unfortunately, we do not know how close the asking and transaction prices are. For the French new car market, \cite{dh18} find that car buyers obtain an average discount of 10\% from the list price. Furthermore, they find considerable heterogeneity in the discount rate with respect to buyer and car characteristics.

\begin{figure}[h!]
\caption{Average first asking price of used diesel vehicles by car make.} \label{fig4}
\centering
\fbox{\includegraphics[width=0.6\textwidth]{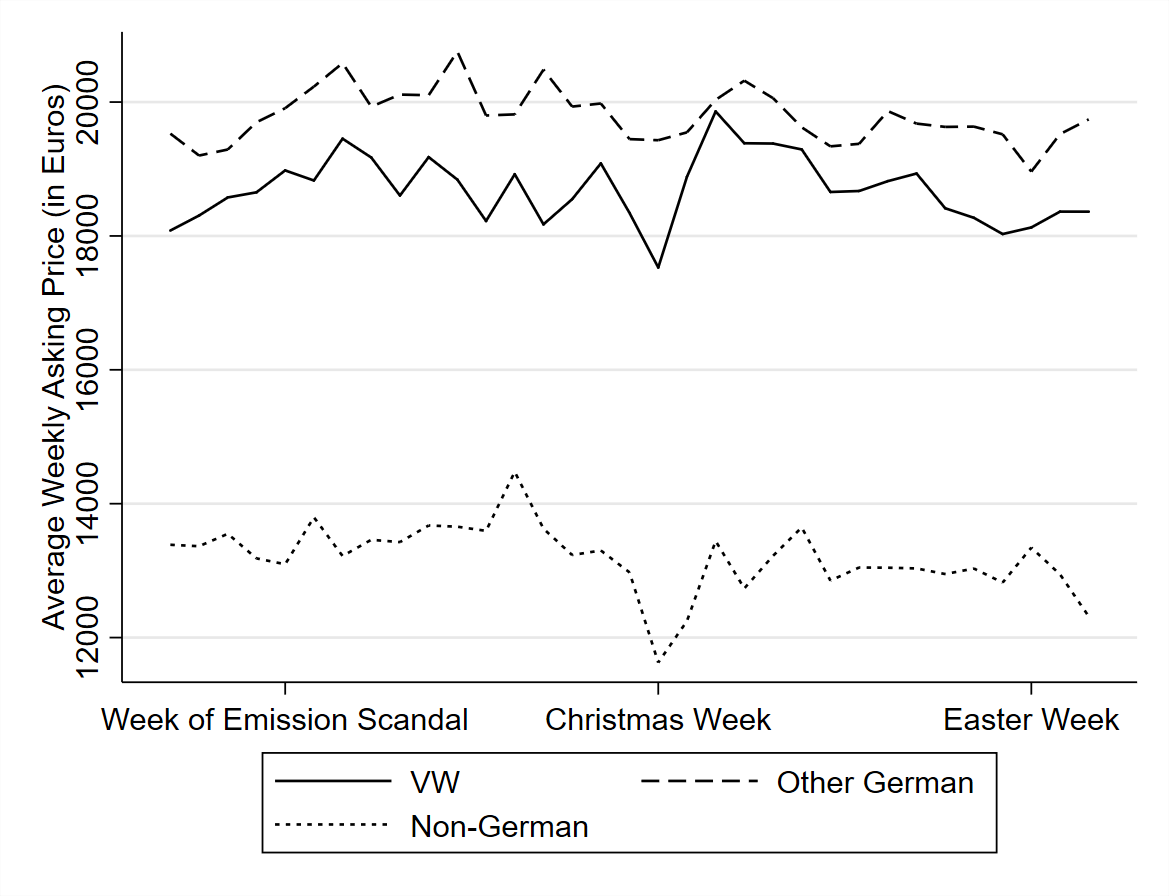} }
\parbox{\textwidth}{\footnotesize \emph{Notes:} This figure plots the average weekly first asking price of used diesel vehicles by car make. Other German car makes include Mercedes, BMW, Opel, and Ford. Non-German car makes include Renault, Peugeot, Fiat, and Toyota. }
\end{figure}

\begin{figure}[h!]
\caption{Average first asking price of used gasoline vehicles by car make.} \label{fig5}
\centering
\fbox{\includegraphics[width=0.6\textwidth]{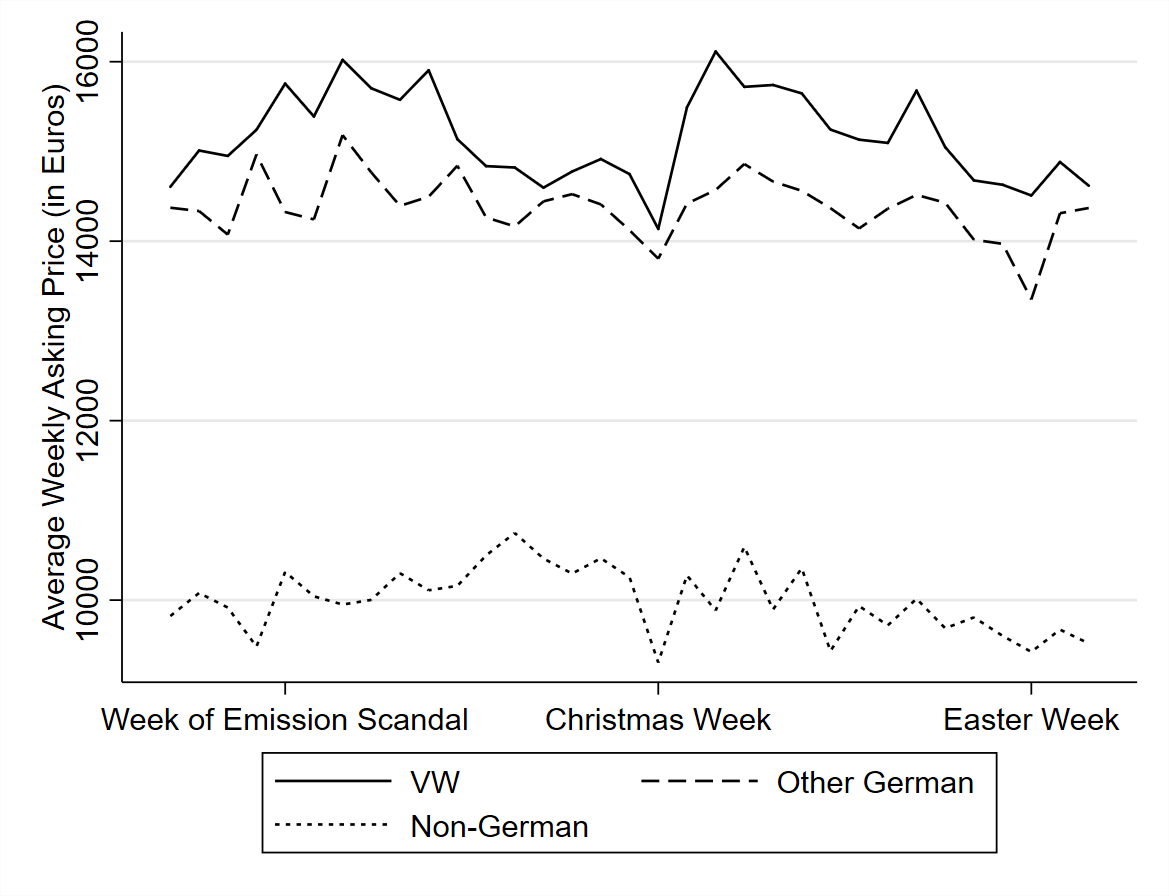} }
\parbox{\textwidth}{\footnotesize \emph{Notes:} This figure plots the average weekly first asking price of used gasoline vehicles by car make. Other German car makes include Mercedes, BMW, Opel, and Ford. Non-German car makes include Renault, Peugeot, Fiat, and Toyota. }
\end{figure}

Figure \ref{fig4} shows the weekly average first asking prices of diesel cars by car make. The asking price is much higher for German than non-German car makes. We observe a price dip during the Christmas week. Other than that, the average first asking prices are remarkably stable over time. Figure \ref{fig5} shows a similar pattern for gasoline cars.

\[ \text{Figures \ref{fig4} and \ref{fig5} around here}  \]

\begin{figure}[h!]
\caption{Kaplan-Meier survival curve of VW diesel vehicles.} \label{fig6}
\centering
\fbox{\includegraphics[width=0.6\textwidth]{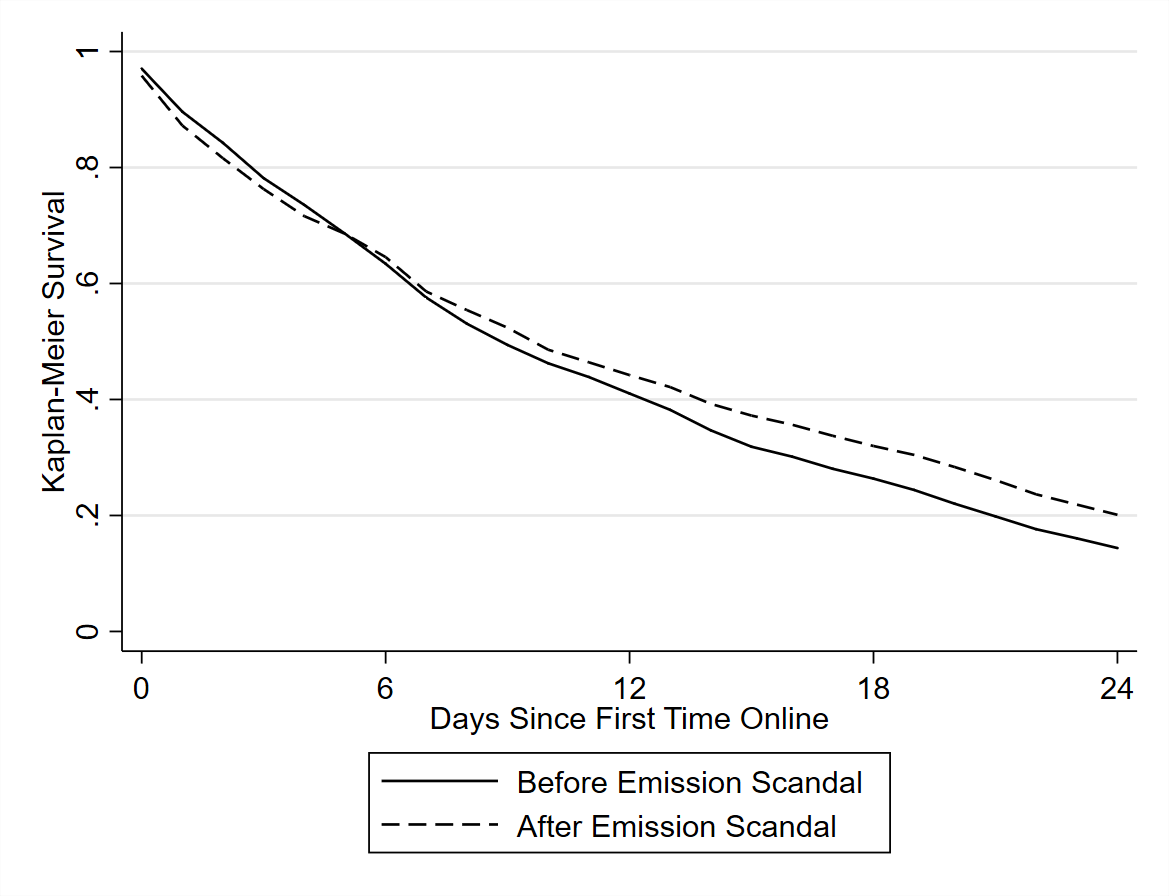} }
\parbox{\textwidth}{\footnotesize \emph{Notes:} This figure plots Kaplan-Meier survival rates of VW diesel vehicles before and after the disclosure of the VW emission scandal. Survival rates report the share of vehicles remaining on the used car market for a certain duration. }
\end{figure}

\begin{figure}[h!]
\caption{Kaplan-Meier survival curve of VW gasoline vehicles.} \label{fig7}
\centering
\fbox{\includegraphics[width=0.6\textwidth]{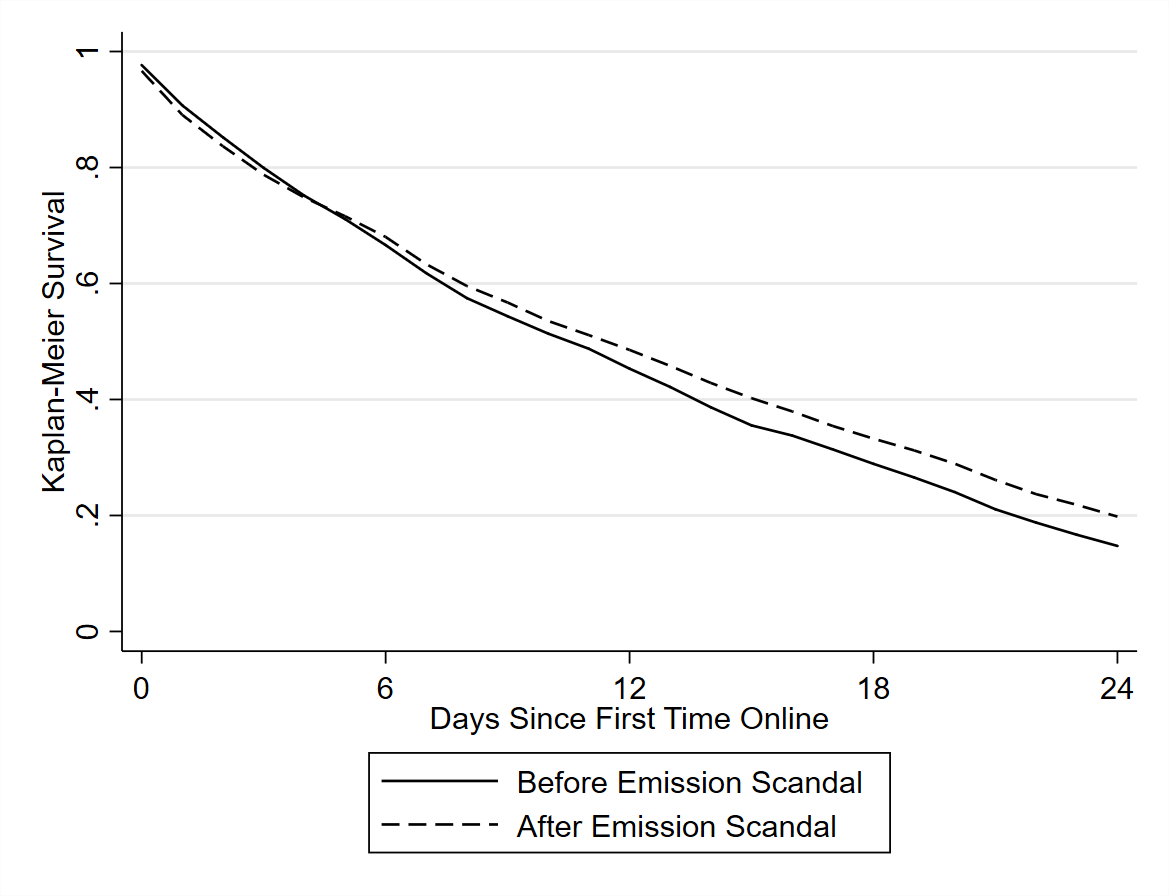} }
\parbox{\textwidth}{\footnotesize \emph{Notes:} This figure plots Kaplan-Meier survival rates of VW gasoline vehicles before and after the disclosure of the VW emission scandal. Survival rates report the share of vehicles remaining on the used car market for a certain duration. }
\end{figure}

\begin{figure}[h!]
\caption{Kaplan-Meier survival curve of non-VW diesel vehicles.} \label{fig8}
\centering
\fbox{\includegraphics[width=0.6\textwidth]{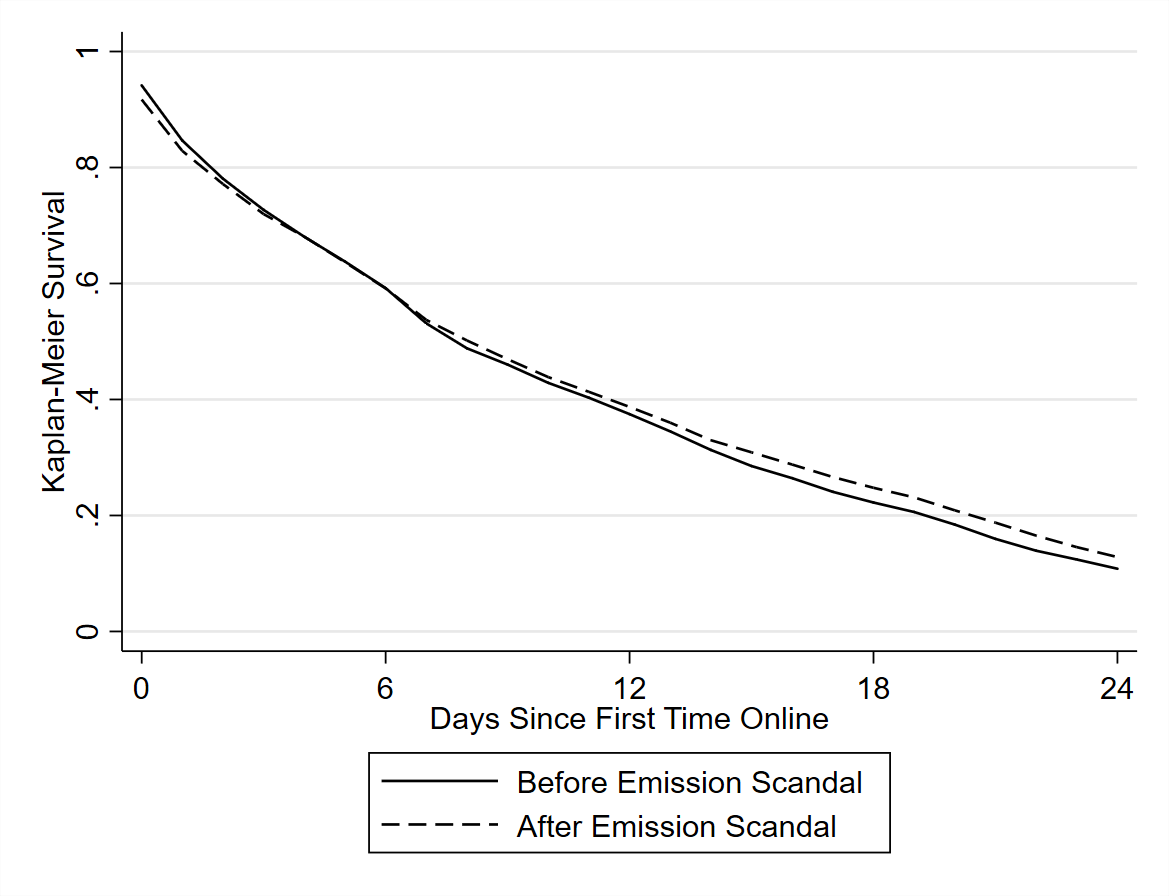} }
\parbox{\textwidth}{\footnotesize \emph{Notes:} This figure plots Kaplan-Meier survival rates of non-VW diesel vehicles before and after the disclosure of the VW emission scandal. Survival rates report the share of vehicles remaining on the used car market for a certain duration. Non-VW vehicles include Mercedes, BMW, Opel, Ford, Renault, Peugeot, Fiat, and Toyota cars. }
\end{figure}

\begin{figure}[h!]
\caption{Kaplan-Meier survival curve of non-VW gasoline vehicles.} \label{fig9}
\centering
\fbox{\includegraphics[width=0.6\textwidth]{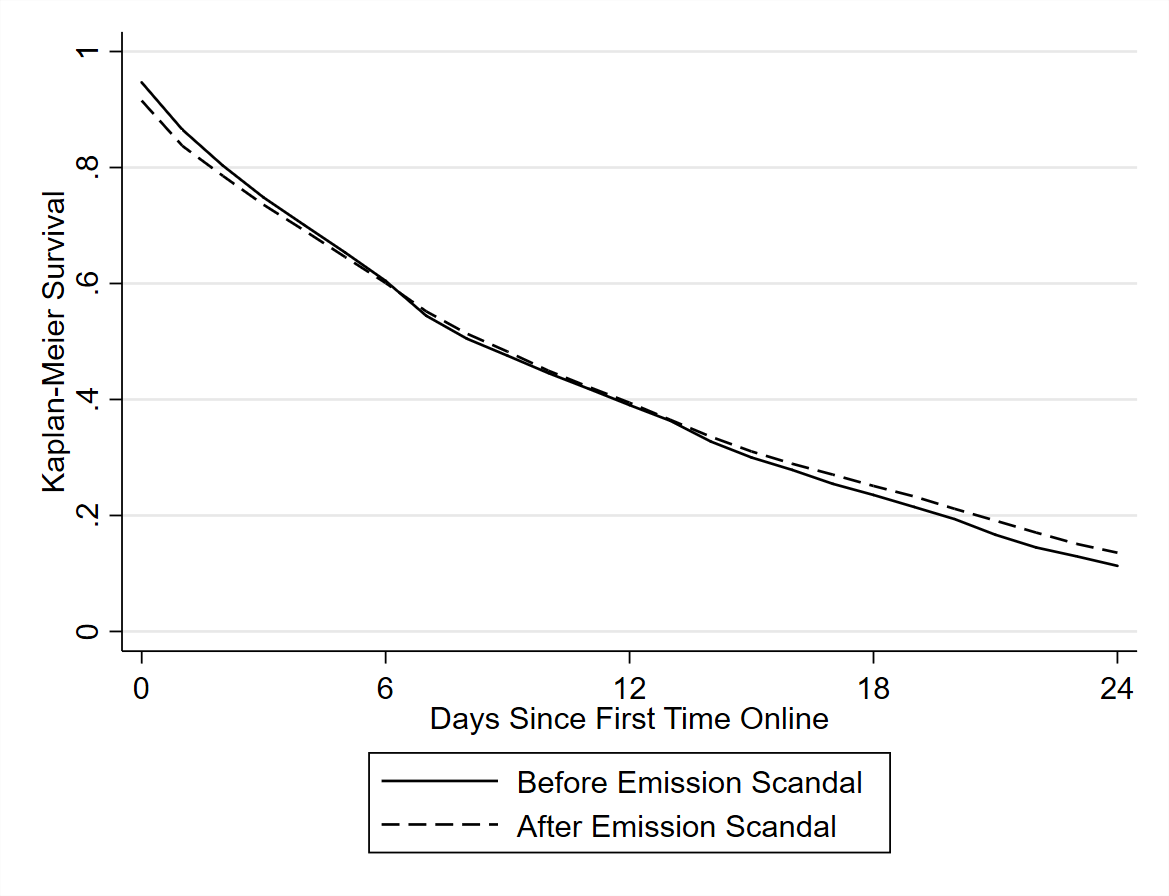} }
\parbox{\textwidth}{\footnotesize \emph{Notes:} This figure plots Kaplan-Meier survival rates of non-VW gasoline vehicles before and after the disclosure of the VW emission scandal. Survival rates report the share of vehicles remaining on the used car market for a certain duration. Non-VW vehicles include Mercedes, BMW, Opel, Ford, Renault, Peugeot, Fiat, and Toyota cars. }
\end{figure}

We calculate the duration between the first and last day online. We observe that more than 95\% of the car advertisements are no longer online after a period of 4 months. However, we observe cars for at most 24 days before the disclosure of the VW emission scandal. The online duration is contaminated when the car advertisement was uploaded before the disclosure of the VW emission scandal and deleted afterwards. To account for this, we calculate Kaplan-Meier survival rates for the first 24 days online. Figures \ref{fig6}-\ref{fig9} show that the online duration is slightly longer after than before the disclosure of the VW emission scandal. This difference is somewhat more pronounced for VW than non-VW cars.

\[ \text{Figures \ref{fig6}, \ref{fig7}, \ref{fig8}, and \ref{fig9} around here}  \]

\section{Empirical approach \label{sec5}}

We compare the share and asking prices of diesel cars before and after the disclosure of the emission scandal and between VW and the other carmakers.\footnote{\rl The expected value of the diesel dummy equals the share of diesel cars. Accordingly, we can use the individual car data to calculate the diesel share. We do not use the count of inflowing diesel cars as an outcome variable because this would imply that we would have to aggregate the data over some periods (e.g., days or weeks). This would result in a loss of information about the individual car characteristics. We do not estimate the effects on the share of VW diesel cars among all diesel cars because doing so would necessitate merging the diesel cars from all brands. In this case, we would lose the control group.} We consider the share and not the quantity of diesel cars because general market conditions affect the share less (e.g., the used car market shows strong seasonality). The dummy variable $D$, $d \in {0,1}$, defines diesel ($D=1$) and gasoline cars ($D=0$). The variable $Y$ is the first asking price (in euros). We distinguish eight time periods in our analyses, $T$, $t \in {0,1,2,3,4,5,6,7}$. The first time period $t=0$ is just before the VW emission scandal was disclosed. The other periods indicate the first seven months after the disclosure. Furthermore, we distinguished between two groups, $G$, $g \in {0,1}$. The first group includes VW cars ($g=1$), and the second group includes cars from other carmakers ($g=0$). The share of diesel vehicles by group and time is indicated by $E[D|G=g,T=t]$. We estimate the effect of the disclosure of the VW emission sandal on the share of VW diesel cars by
\begin{align*}
\delta_s=&(E[D|G=1,T=s]-E[D|G=1,T=0])\\&-(E[D|G=0,T=s]-E[D|G=0,T=0]),
\end{align*}
for $s \in {1,2,3,4,5,6,7}$. We estimate the effects on the first asking price of VW diesel cars by
\begin{align*}
\alpha_s=&(E[Y|G=1,T=s,D=1]-E[Y|G=1,T=0,D=1])\\ &-(E[Y|G=0,T=s,D=1]-E[Y|G=0,T=0,D=1]).
\end{align*}
We assume that the decline in observable quality is stronger for VW diesel cars than for cars from other carmakers. However, $\delta_s$ and $\alpha_s$ are conservative estimates because other carmakers could also be affected by the disclosure of the emission scandal. If the VW diesel scandal affects the share of both groups in the same direction, then we underestimate the true effect. Therefore, $\delta_s$ and $\alpha_s$ could be lower bounds of the impacts of the emission scandal on the share and asking price of VW diesel cars. Alternatively, if some individuals have a particular preference for diesel cars, the emissions scandal might reduce the attractiveness of VW diesels cars but increase demand for non-VW diesel cars. This increase in demand could affect the supply and price of used cars and lead to a positive bias. In Section \ref{sec63}, we provide some evidence against this alternative hypothesis of a positive bias by using an alternative control group.

It could be that observable car characteristics change between the groups over time. For example, sellers could bring older cars to market. Because diesel cars tend to have a longer life-cycle than gasoline cars, this type of market behavior would increase the share of diesel cars. To account for this, we define the conditional share of diesel cars $E[D|G=g,T=t,X=x]$, where $X$ contains observable car and seller characteristics (see Table B.1 in Online Appendix B for a list of all control variables we consider). We estimate the effect of the disclosure of the emission scandal on the share of VW cars with diesel engines conditional on observable characteristics.
\begin{align*}
\gamma_s (x)=&(E[D|G=1,T=s,X=x]-E[D|G=1,T=0,X=x])\\ &-(E[D|G=0,T=s,X=x]-E[D|G=0,T=0,X=x]) \mbox{ and}\\
\mu_s (x)= &(E[Y|G=1,T=s,D=1,X=x]-E[Y|G=1,T=0,D=1,X=x])\\ &-(E[Y|G=0,T=s,D=1,X=x]-E[Y|G=0,T=0,D=1,X=x]).
\end{align*}
The parameters $\gamma_s (x)$ and $\mu_s (x)$ measure the effects of the emission scandal's disclosure on the share and asking price of VW cars with diesel engines for cars and sellers with similar observable characteristics. We integrate the conditional effects to
\begin{equation*}
\gamma_s =E_{X|G=1,T=0} [\gamma_s (x)|G=1,T=0] \mbox{ and }
\mu_s  =E_{X|G=1,T=0,D=1} [\mu_s (x)|G=1,T=0,D=1],
\end{equation*}
which are the average effects defined for cars and sellers with characteristics that are similar to the VW group in the pre-disclosure period. Intuitively, this is the direct effect of the VW emission scandal after controlling for other possible market distortions. If we find an effect of the emission scandal on the share or asking price of diesel cars, even after controlling for a large set of important observable characteristics, then this is a strong indication of sorting.

We perform estimations in separate samples for each group and each period. We estimate the conditional moments with radius matching on the propensity score \citep[see][]{le11}.\footnote{\rl The (main) advantage of matching as opposed to (widely used) linear regression-based difference-in-differences approaches is the robustness to misspecification of the linear model.} The propensity scores
are parametrically specified by Probit models, and the conditional expectations of the outcomes are unrestricted and thus non-parametric. The algorithm is more precise than nearest-neighbor matching due to the idea of radius matching \citep[e.g.,][]{de02}. Furthermore, the procedure uses the initial matching weights for a (weighted) regression adjustment for bias reduction in a second step \citep[see][]{ab11}.\footnote{\rl Therefore, the estimator satisfies the so-called double robustness property, implying that it is consistent if either the propensity score or the regression model is correctly specified \citep[e.g.,][]{ru79,jo04}.} This regression adjustment should also reduce small sample and asymptotic biases of matching. \cite{hu13} investigate the finite sample properties of this algorithm along with other matching type estimators and find it to be very competitive.

We set the radius size to 90\% of the 0.9th quantile of the distance between matched treated and control observations occurring in standard nearest-neighbor matching. Inference is based on bootstrapping the respective effect 499 times and using the standard deviation of the bootstrapped effects as an estimate of the standard error of the t-statistic. \cite{ab08} show that bootstrap-based standard errors may be invalid for matching based on a fixed number of comparison observations. However, our matching algorithm is smoother than the latter approach because it (by the nature of radius matching) uses a variable number of comparisons that are distance-weighted within the radius and, moreover, applies the regression adjustment. Therefore, the bootstrap is likely to be a valid inference procedure for the radius matching estimator used here. It performs well in a large-scale (empirically based) simulation study by \cite{bo16}, who investigate the performance of several variance estimators in the context of propensity score-based matching estimation.

\section{Results \label{sec6}}
\subsection{Unconditional estimates}

\begin{figure}[h!]
\caption{Difference in the share of used diesel vehicles.} \label{fig10}
\centering
\fbox{\includegraphics[width=0.6\textwidth]{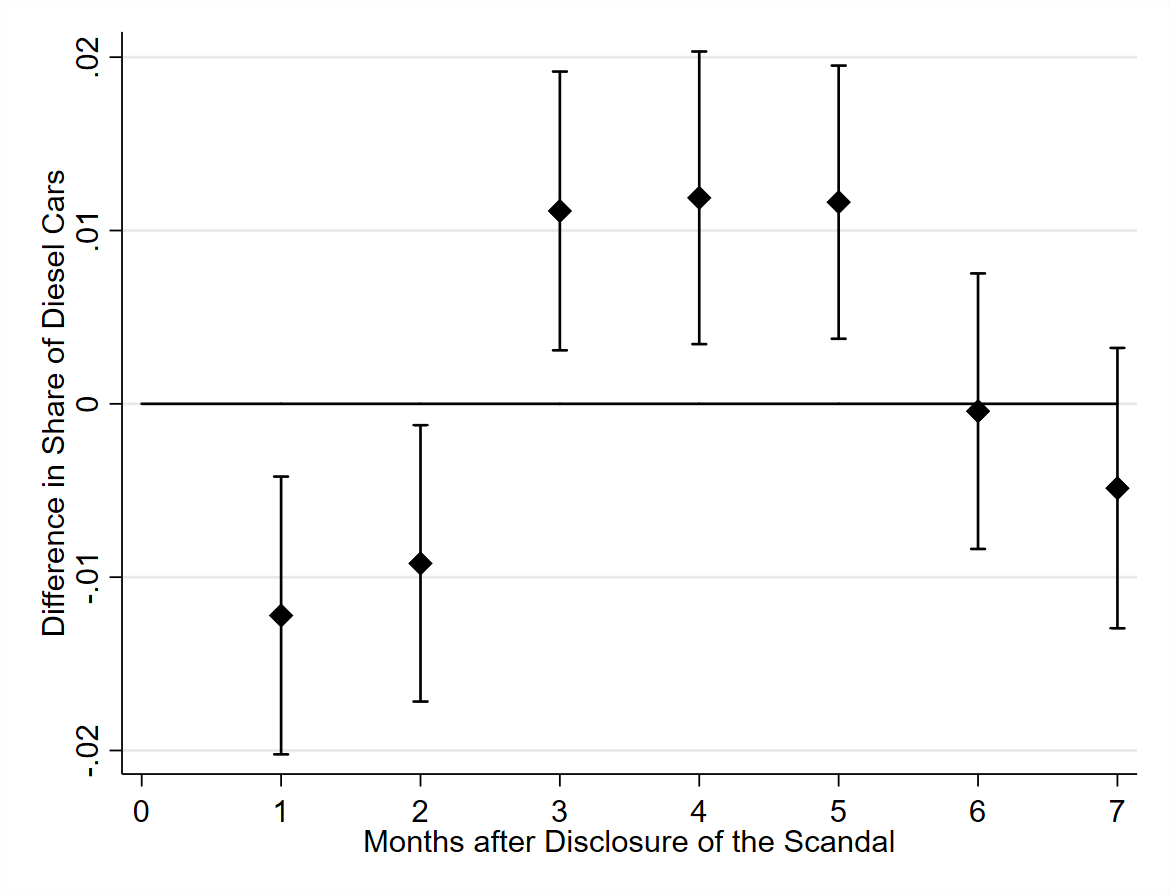} }
\parbox{\textwidth}{\footnotesize \emph{Notes:} This figure plots the difference-in-differences of the used car diesel share for the first seven months after the disclosure of the VW emission scandal. The first difference is between the time before and after the disclosure of the VW emission scandal. The second difference is between VW and other car makes. The diamonds indicate the resulting differences in the share of used diesel vehicles. The capped vertical lines indicate the 95\% confidence intervals of the monthly averages. The confidence intervals are calculated with a non-parametric bootstrap (499 replications). The other car makes include Mercedes, BMW, Opel, Ford, Renault, Peugeot, Fiat, and Toyota cars. }
\end{figure}

Figure \ref{fig10} documents the unconditional estimates of supply. They reflect the difference in the share of vehicles with diesel engines between VW and other carmakers. During the first two months after the disclosure, we find a one-percentage-point lower share of VW diesel cars; between months three and five, we find a one-percentage-point higher share of VW diesel cars. In months six and seven, the difference in the shares does not differ significantly from zero.

\[ \text{Figure \ref{fig10} around here}  \]

The results could reflect delay and catch-up effects. This time pattern could be due to an initial shock of the scandal reducing the supply (possibly a response to a decrease in demand). Once the news wears off and demand returns to levels similar to previous ones, all the cars that were not supplied immediately after the scandal join the cars that would have been supplied in any case, leading to an increase in supply. Then, once the `delayed supply' is sold off, supply returns to normal levels. However, sellers can advertise their cars even when there is little demand. Furthermore, the VW scandal was not settled within six months, and it persists to the present.

An alternative explanation for this pattern is that car owners retrofitted their manipulated cars before selling them. However, this explanation is unlikely because the retrofitting plans were submitted approximately 1.5 months after the revelation of the emission scandal for approval to the German Federal Motor Transport Authority. It followed a long-lasting political debate about the optimal retrofitting plan. To date, many manipulated cars have not received retrofitting in Germany.

\begin{figure}[h!]
\caption{Difference in the first asking price of diesel vehicles.} \label{fig11}
\centering
\fbox{\includegraphics[width=0.6\textwidth]{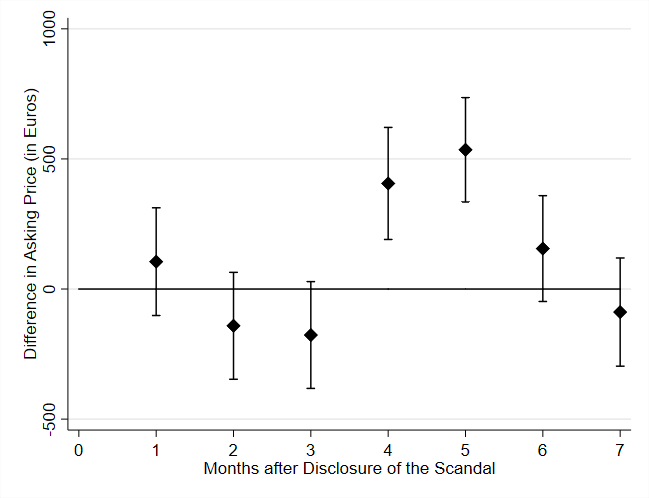} }
\parbox{\textwidth}{\footnotesize \emph{Notes:} This figure plots the difference-in-differences in the first asking price of used diesel vehicles for the first seven months after the disclosure of the VW emission scandal. The first difference is between the time before and after the disclosure of the VW emission scandal. The second difference is between VW and other car makes. The diamonds indicate the resulting differences in the first asking price of used diesel vehicles. The capped vertical lines indicate the 95\% confidence intervals of the monthly averages. The confidence intervals are calculated with a non-parametric bootstrap (499 replications). The other car makes include Mercedes, BMW, Opel, Ford, Renault, Peugeot, Fiat, and Toyota diesel cars.}
\end{figure}

Figure \ref{fig11} shows the estimates for first asking prices. The differences are mostly insignificant, but we find positive asking price differences four and five months after the disclosure of the scandal. The change in the asking prices of VW diesel cars is up to 500 euros higher than the change in non-VW diesel cars. It is possible that VW diesel cars with better quality (e.g., lower mileage) came to the market after the disclosure of the VW emission scandal. This could lead to an upward bias of the asking price estimates if we do not control for important car characteristics. We discuss the conditional effects in the next section.

\[ \text{Figure \ref{fig11} around here}  \]

\subsection{Conditional estimates}

\begin{figure}[h!]
\caption{Matched difference in the share of used diesel vehicles.} \label{fig12}
\centering
\fbox{\includegraphics[width=0.6\textwidth]{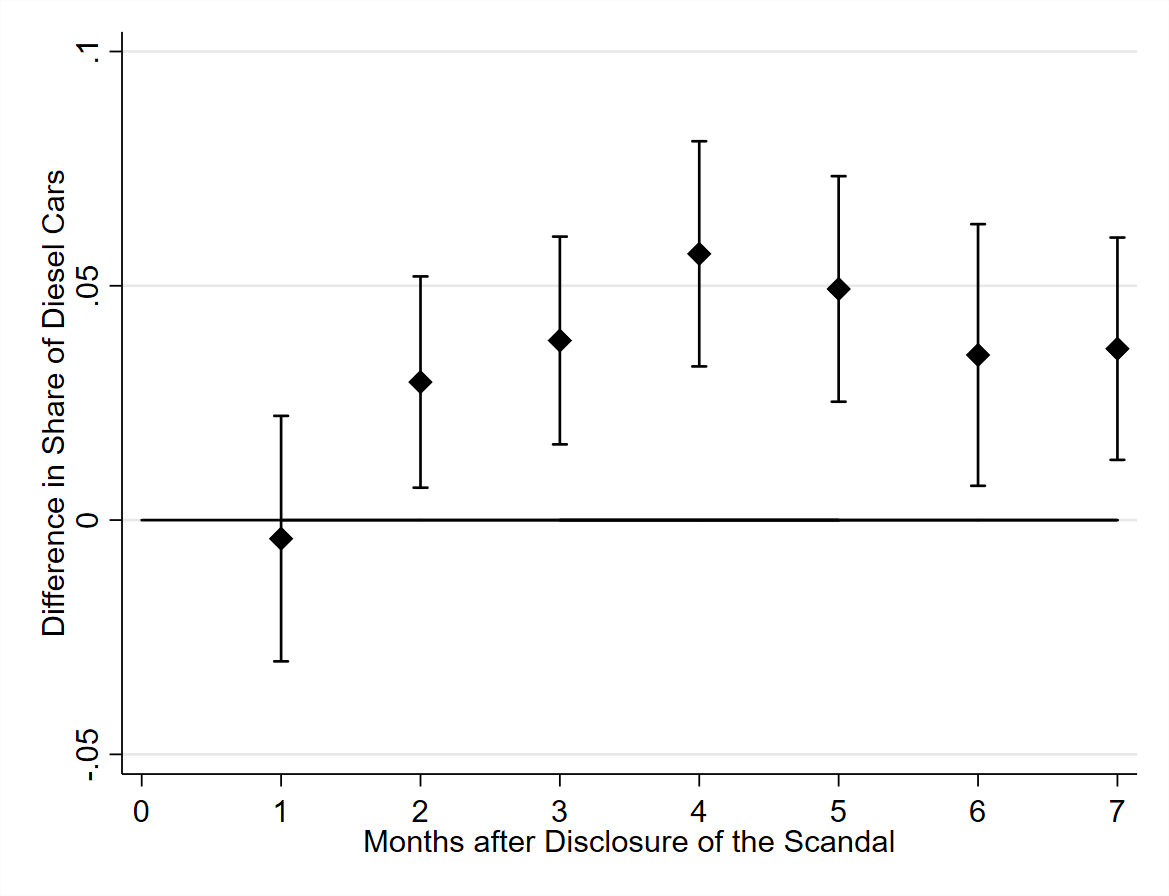} }
\parbox{\textwidth}{\footnotesize \emph{Notes:} This figure plots the matched difference-in-differences of the used car diesel share for the first seven months after the disclosure of the VW emission scandal. Diesel and gasoline cars are matched at the individual car level such that they share the same characteristics. The first matched difference is between the time before and after the disclosure of the VW emission scandal. The second matched difference is between VW and other car makes. The diamonds indicate the resulting matched differences in the share of used diesel vehicles. The capped vertical lines indicate the 95\% confidence intervals of the monthly averages. The confidence intervals are calculated with a non-parametric bootstrap (499 replications). The other car makes include Mercedes, BMW, Opel, Ford, Renault, Peugeot, Fiat, and Toyota cars. }
\end{figure}

Figure \ref{fig12} presents the results of the conditional difference-in-differences approach. We control for different dummies for car age and mileage. Following the specification in \cite{la12}, we control for the first seven polynomials of mileage. We include dummies for vehicle class and car body. Additionally, we include dummies for full service history, warranty, the validity of general inspection, and private seller. We control for different emission standards and fuel consumption of the offered car. Tables B.1-B.4 in Online Appendix B present the balance of all control variables that we consider before matching by group and time period. We find large imbalances, especially between the two groups of VW cars (diesel and gasoline engines) and other carmakers. The cars from other makers are, on average, older than VW cars. Additionally, the share of Euro 6 cars and the age of cars within the groups increases over time. This could explain the delay and catch-up effects because, for example, used diesel vehicles are on average older than gasoline vehicles. Furthermore, this could explain the positive asking price estimates for VW diesel vehicles.

\[ \text{Figure \ref{fig12} around here}  \]

We present the marginal effects of the propensity score estimates in Tables C.1-C.2 of Online Appendix C. The propensity score estimates serve as inputs to the matching algorithm. When performing matching, one should check for potential issues of (i) insufficient support in the propensity scores across treatment states that may result in incomparable matches as well as large matching weights of some non-treated observations with specific propensity scores and (ii) imbalances in covariates after matching (due to inappropriate propensity score specifications). To account for support problems, we drop treated observations with propensity score values above the highest propensity score value in the control group. \cite{le19} show that this support procedure can improve the finite sample efficiency of the estimates. We document the distribution of the propensity scores of the observations on support for all samples in Figures D.1-D.15 in Online Appendix D. As a final specification test, we document the balancing of the control variables after matching in Tables E.1-E.4 of Online Appendix E. We find only small imbalances between groups and periods. The standardized differences are always below 10.

The conditional results suggest no significant effects on the share of VW diesel vehicles during the first two months after the disclosure of the emission scandal. Subsequently, we find significant positive effects on the share of VW diesel vehicles between months 3 and 7. In months 4 and 5, the share of VW diesel vehicles increases by 4 percentage points. In months 6 and 7, the effects on the VW diesel share decline but are still positive and significant. Accordingly, the results are robust and even more positive after we control for a large set of relevant covariates. The positive effects clearly dominate the insignificant effects during the first two months. The findings cannot merely reflect delay and catch-up effects.

\begin{figure}[h!]
\caption{Matched difference in the first asking price of diesel vehicles.} \label{fig13}
\centering
\fbox{\includegraphics[width=0.6\textwidth]{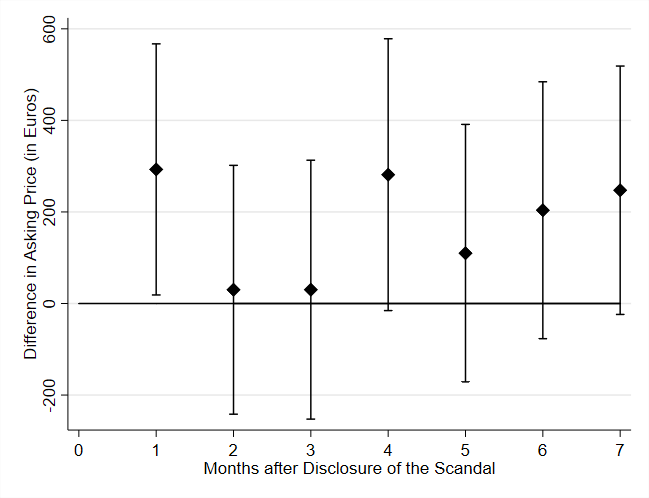} }
\parbox{\textwidth}{\footnotesize \emph{Notes:} This figure plots the matched difference-in-differences in the first asking price of used diesel vehicles for the first seven months after the disclosure of the VW emission scandal. Diesel cars are matched at the individual car level such that they share the same characteristics. The first matched difference is between the time before and after the disclosure of the VW emission scandal. The second matched difference is between VW and other car makes. The diamonds indicate the resulting matched differences in the first asking price of used diesel vehicles. The capped vertical lines indicate the 95\% confidence intervals of the monthly averages. The confidence intervals are calculated with a non-parametric bootstrap (499 replications). The other car makes include Mercedes, BMW, Opel, Ford, Renault, Peugeot, Fiat, and Toyota diesel cars.}
\end{figure}

Figure \ref{fig13} reports the conditional results for first asking prices. The conditional differences have a much smaller magnitude than the unconditional differences and are not significantly different from zero. This suggests that the unconditional price estimates are upward biased because different types of cars came to market after the disclosure of the VW emission scandal.

\[ \text{Figure \ref{fig13} around here}  \]

Table \ref{tab4} documents the effects on the share of VW diesel vehicles by environmental standard. Cars with defeat devices are always regulated under the Euro 5 emission standard. Cars with emission standards below or above Euro 5 have no defeat device. We consider the results for those groups as a placebo sensitivity check. For VW diesel cars with an emission standard less stringent than Euro 5, we do not find any effects. For VW diesel cars regulated under the Euro 6 emission standard, we find negative effects six months after the disclosure. Accordingly, the results for cars without a defeat device are either insignificant or move in the opposite direction. This is evidence in favor of the placebo sensitivity check. In contrast, the share of VW diesel cars regulated under the Euro 5 emission standard significantly increased after the emission scandal, which is consistent with the predictions of the small sorting model.

\[ \text{Table \ref{tab4} around here}  \]

Table \ref{tab5} reports the estimates of the first asking price by emission standard. For VW diesel cars regulated under the Euro 5 emission standard, we find clear evidence for negative asking price effects. Accordingly, the results for the VW diesel cars that have a high probability of manipulation are in line with the predictions of the small sorting model. However, we find positive effects on asking prices of VW diesel cars regulated under the Euro 6 emission standard. These cars have lower emissions than Euro 5 and are supposed to have no defeat device. It appears that the positive and negative estimates of Euro 5 and 6 cars cancel out. This could explain why there are on average no significant differences. Professional car sellers could make Euro 6 cars more expensive after the disclosure of the VW emission scandal to decrease their attractiveness for potential buyers relative to Euro 5 VW diesel cars. The results for cars with an emission standard below Euro 5 are ambiguous.

\[ \text{Table \ref{tab5} around here}  \]

In sum, for cars with a high probability of a defeat device, we find positive supply and negative asking price effects. These findings are consistent with the predictions of the sorting model above. The spillover effects to not-manipulated VW diesel are inconsistent with the predictions of the sorting model.

\subsection{Alternative control group \label{sec63}}

It is challenging to find a suitable control group for difference-in-difference estimators. On the one hand, the control and treatment groups should be similar, such that the two groups have comparable market trends. On the other hand, overly similar control and treatment groups could lead to spillover effects.

\begin{figure}[h!]
\caption{Matched difference in the share of used diesel vehicles between VW and non-German car makes.} \label{fig14}
\centering
\fbox{\includegraphics[width=0.6\textwidth]{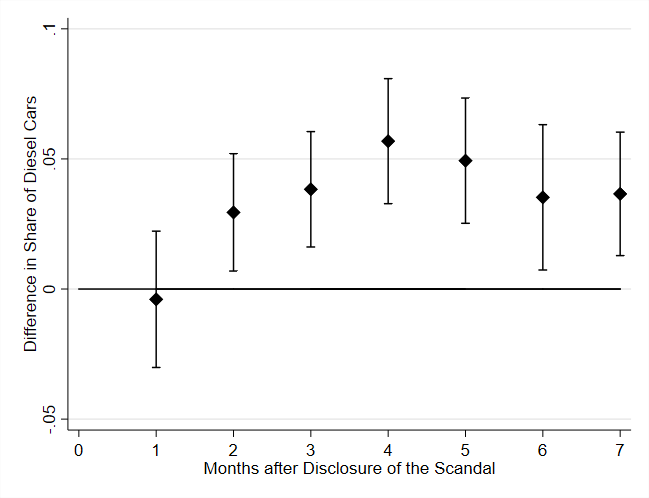} }
\parbox{\textwidth}{\footnotesize \emph{Notes:} This figure plots the matched difference-in-differences of the used car diesel share for the first seven months after the disclosure of the VW emission scandal. Diesel and gasoline cars are matched at the individual car level such that they share the same characteristics. The first matched difference is between the time before and after the disclosure of the VW emission scandal. The second matched difference is between VW and non-German car makes. The diamonds indicate the resulting matched differences in the share of used diesel vehicles. The capped vertical lines indicate the 95\% confidence intervals of the monthly averages. The confidence intervals are calculated with a non-parametric bootstrap (499 replications). The non-German car makes include Renault, Peugeot, Fiat, and Toyota cars. }
\end{figure}

\begin{figure}[h!]
\caption{Matched difference in the first asking price of diesel vehicles between VW and non-German car makes.} \label{fig15}
\centering
\fbox{\includegraphics[width=0.6\textwidth]{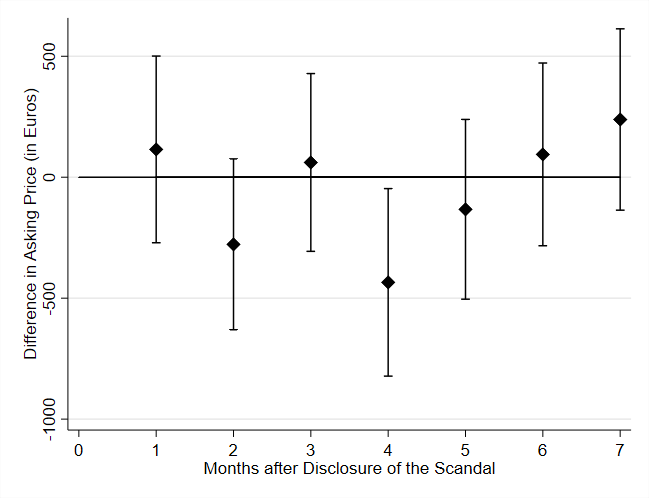} }
\parbox{\textwidth}{\footnotesize \emph{Notes:} This figure plots the matched difference-in-differences in the first asking price of used diesel vehicles for the first seven months after the disclosure of the VW emission scandal. Diesel cars are matched at the individual car level such that they share the same characteristics. The first matched difference is between the time before and after the disclosure of the VW emission scandal. The second matched difference is between VW and non-German car makes. The diamonds indicate the resulting matched differences in the first asking price of used diesel vehicles. The capped vertical lines indicate the 95\% confidence intervals of the monthly averages. The confidence intervals are calculated with a non-parametric bootstrap (499 replications). The non-German car makes include Renault, Peugeot, Fiat, and Toyota diesel cars.}
\end{figure}

Figures \ref{fig14} and \ref{fig15} report the findings of a sensitivity test using only non-German car makes in the control group. Arguably, non-German car makes are less affected by the disclosure of the VW emission scandal than German car makes, but they might follow different market trends. Qualitatively, the results are similar regardless of whether we include German cars in the control group. However, quantitatively, the effects are slightly larger when we exclude German car makes, which is supporting evidence of a negative bias. Furthermore, we find significant negative asking price estimates in the fourth month after the disclosure. The findings of this sensitivity check suggest that the magnitudes of the main results are a lower bound.

\[ \text{Figures \ref{fig14} and \ref{fig15} around here}  \]

\subsection{Effect heterogeneity}

Table \ref{tab6} shows the effect of the VW emission scandal on the share of diesel cars by vehicle class. The positive effects are mainly driven by minivans and SUVs. These cars are considered heavy polluters because of their size and weight. Accordingly, the empirical results are coherent for those vehicle classes that are most affected by the disclosure of the VW emission scandal. Moreover, we also find significant positive effects on the share of diesel cars for compact and medium-sized cars. Small cars are the only class for which we do not find a significant increase in the share of diesel cars. If anything, we observe a negative delayed effect in the first month after the disclosure of the emission manipulation scandal. Table \ref{tab7} reports that compact cars (VW Golf) are mainly responsible for the negative first asking price differences.

\[ \text{Tables \ref{tab6} and \ref{tab7} around here}  \]

Table \ref{tab8} reports effect heterogeneity by seller type. The share of VW diesel cars does not increase among private sellers. In contrast, the share of VW diesel cars actually declines among private sellers. For professional car dealers, we observe a significant increase in the share of VW diesel cars. Only in the first month after the disclosure did we find a negative delayed effect. It is possible that VW diesel car owners more often engage professional car dealers as sales agents, or leasing contracts are terminated instead of extended (i.e., exploiting buyback options). The latter could occur because car lessors believe that the buyback price specified in the leasing contract is higher than the market price after the disclosure. Furthermore, car owners can trade-in their used car when they buy a new car. New car buyers often have a higher preference for quality than used car buyers. Accordingly, this explanation would be consistent with the assumption that car owners with high quality preferences in particular supply their cars.

\[ \text{Table \ref{tab8} around here}  \]

Some professional car dealers provide warranties on offered used cars. The lower part of Table \ref{tab8} shows the effect heterogeneity between professional car dealers with and without an offered warranty. For both groups, we find evidence of an increase in the share of VW diesel vehicles. However, for professional car dealers without a warranty, the positive effect is much more persistent across the sample period, whereas we only find significant positive effects in the fifth month after disclosure for professional car dealers with a warranty. Accordingly, professional car dealers who do not offer warranties are mainly responsible for the increase in the share of VW diesel cars after the emission scandal. It appears that dealers who offer warranties are less willing to offer these warranties on vehicles that are known to have defects in the exhaust system.

Table \ref{tab9} reports effect heterogeneity in first asking prices by seller type. However, the results are ambiguous, and no clear pattern can be detected. If anything, it appears that professional sellers are more willing to offer a price discount than private sellers.

\[ \text{Table \ref{tab9} around here}  \]

\section{Discussion of alternative channels \label{sec7}}

Although the positive supply and negative asking price effects are consistent with the predictions of the small sorting model introduced in Section \ref{sec3}, alternative mechanisms could explain similar empirical patterns. First, the quality of all diesel cars could decline after the diesel scandal because all diesel cars lose reputation. This is certainly the case, but we believe that the quality loss is stronger for VW diesel cars than for other diesel and gasoline cars. The share of non-manipulated VW diesel cars is not altered by the scandal (see Table \ref{tab4}), which is evidence against strong spillover effects. Nevertheless, our results may provide a lower bound for the effect of the VW emission scandal on the supply of used cars. The results of Section \ref{sec63} provide suggestive evidence that spillover effects would bias our results towards zero.

Second, information about the VW scandal could be available to the car owner but may (partially) diffuse among potential buyers over time. This could incentivize owners of manipulated cars to sell early to skim naive buyers. However, the VW diesel scandal received so much media attention that any potential buyer should have been immediately aware of it. It is more likely that car owners of non-manipulated cars attempted to sell before the possible revelation of further manipulations. However, we do not find such a pattern in the data (see Table \ref{tab4}).

Third, risk aversion could be an alternative motive to sorting for supplying manipulated cars. Owners who are risk averse might not want to take a risk in holding on to their manipulated car, and less risk-averse buyers might be glad to accept the risk. Possibly, they do not trust the retrofitting plans or fear future fraud revelations. In the latter case, we would expect to find supply effects for all VW diesel cars and not only for those with a high probability of manipulation. We cannot find evidence for this in the data (see Table \ref{tab4}).

\section{Conclusions \label{sec8}}

We exploit the quasi-experimental decline in the observable environmental quality of VW diesel vehicles after the disclosure of the VW emission manipulation scandal by using a conditional difference-in-differences method based on semi-parametric matching estimators to analyze German car supply around the scandal. We document supply-side reactions to the environmental quality of cars. The supply of VW cars with a high probability of having a defeat device increases after the disclosure of the VW emission manipulation scandal. Furthermore, we find some evidence of negative effects of the VW emission scandal on the asking prices of cars with a high manipulation probability.

We rationalize our findings with the predictions of a simple sorting model. Individuals with high preferences for environmental quality could sell their VW diesel cars to individuals with low preferences for environmental quality. This could increase the market volume even when prices are falling.

The main takeaway of this study is that information about the environmental quality of cars is important for the car market. This suggests that there is demand for better certification processes for the environmental quality of cars.

A disadvantage of our study is that we only observe advertisements for and not transactions of cars. Further research could investigate demand-side reactions to the disclosure of environmental quality information.

\bibliographystyle{ecca}
\bibliography{Bibliothek}

\begin{thebibliography}{28}
\providecommand{\natexlab}[1]{#1}

\bibitem[{Abadie and Imbens(2008)}]{ab08}
\textsc{Abadie, A.} and \textsc{Imbens, G.} (2008). {On the Failure of the
  Bootstrap for Matching Estimators}. \textit{Econometrica}, \textbf{76}~(6),
  1537–1557.

\bibitem[{Abadie and Imbens(2011)}]{ab11}
\textsc{---} and \textsc{---} (2011). {Bias-Corrected Matching Estimators for
  Average Treatment Effects}. \textit{Journal of Business and Economic
  Statistics}, \textbf{29}~(1), 1--11.

\bibitem[{Allcott and Wosny(2013)}]{al13}
\textsc{Allcott, H.} and \textsc{Wosny, N.} (2013). {Gasoline Prices, Fuel
  Economy, and the Energy Paradox}. \textit{Review of Economics and
  Statistics}, \textbf{96}~(5), 779--795.

\bibitem[{Bodory \textit{et~al.}(2016)Bodory, Camponovo, Huber and
  Lechner}]{bo16}
\textsc{Bodory, H.}, \textsc{Camponovo, L.}, \textsc{Huber, M.} and
  \textsc{Lechner, M.} (2016). {The Finite Sample Performance of Inference
  Methods for Propensity Score Matching and Weighting Estimators}.
  \textit{Journal of Business and Economic Statistics}, \textbf{forthcoming}.

\bibitem[{Busse \textit{et~al.}(2013)Busse, Knittel and Zettelmeyer}]{bu13}
\textsc{Busse, M.}, \textsc{Knittel, C.} and \textsc{Zettelmeyer, F.} (2013).
  {Are Consumers Myopic? Evidence from New and Used Car Purchases}.
  \textit{American Economic Review}, \textbf{103}~(1), 220--256.

\bibitem[{Clemente and Gabbioneta(2017)}]{cl17}
\textsc{Clemente, M.} and \textsc{Gabbioneta, C.} (2017). {How Does the Media
  Frame Corporate Scandals? The Case of German Newspapers and the Volkswagen
  Diesel Scandal}. \textit{Journal of Management Inquiry},
  \textbf{forthcoming}.

\bibitem[{Dehejia and Wahba(2002)}]{de02}
\textsc{Dehejia, R.} and \textsc{Wahba, S.} (2002). {Propensity-Score-Matching
  Methods for Nonexperimental Causal Studies}. \textit{Review of Economics and
  Statistics}, \textbf{84}~(1), 151–161.

\bibitem[{D'Hautefoeuille \textit{et~al.}(2018)D'Hautefoeuille, Durrmeyer and
  F\'{e}vier}]{dh18}
\textsc{D'Hautefoeuille, X.}, \textsc{Durrmeyer, I.} and \textsc{F\'{e}vier,
  P.} (2018). {Automobile Prices in Market Equilibrium with Unobserved Price
  Discrimination}. \textit{Review of Economic Studies}, \textbf{forthcoming}.

\bibitem[{EPA(2015)}]{ep15}
\textsc{EPA} (2015). {Notice of Violation}. \textit{sent by EPA to Volkswagen
  Group of America, Inc (18 September 2015)}.

\bibitem[{Hendel and Lizzeri(1999)}]{he99}
\textsc{Hendel, I.} and \textsc{Lizzeri, A.} (1999). {Adverse Selection in
  Durable Goods Markets}. \textit{American Economic Review}, \textbf{89}~(5),
  1097--1115.

\bibitem[{Hendel and Lizzeri(2002)}]{he02}
\textsc{---} and \textsc{---} (2002). {The Role of Leasing Under Adverse
  Selection}. \textit{Journal of Political Economy}, \textbf{110}~(1),
  113--143.

\bibitem[{Huber \textit{et~al.}(2013)Huber, Lechner and Wunsch}]{hu13}
\textsc{Huber, M.}, \textsc{Lechner, M.} and \textsc{Wunsch, C.} (2013). {The
  Performance of Estimators Based on the Propensity Score}. \textit{Journal of
  Econometrics}, \textbf{175}~(1), 1--21.

\bibitem[{Joffe \textit{et~al.}(2004)Joffe, Ten~Have, Feldman and
  Kimmel}]{jo04}
\textsc{Joffe, M.}, \textsc{Ten~Have, T.}, \textsc{Feldman, H.} and
  \textsc{Kimmel, S.} (2004). {Model Selection, Confounder Control, and
  Marginal Structural Models: Review and New Applications}. \textit{American
  Statistician}, \textbf{58}~(4), 272–279.

\bibitem[{Kraftfahrtbundesamt(2017)}]{kr17}
\textsc{Kraftfahrtbundesamt} (2017). {Fahrzeugzulassungen (FZ) Neuzulassungen,
  Besitzumschreibungen, Ausserbetriebsetzung von Kraftfahrzeugen und
  Kraftfahrzeuganh\"{a}ngern}. \textit{FZ 7}.

\bibitem[{Lacetera \textit{et~al.}(2012)Lacetera, Pope and Sydnor}]{la12}
\textsc{Lacetera, N.}, \textsc{Pope, D.} and \textsc{Sydnor, H.} (2012).
  {Heuristic Thinking and Limited Attention in the Car Market}.
  \textit{American Economic Review}, \textbf{102}~(5), 2206–2236.

\bibitem[{Lechner \textit{et~al.}(2011)Lechner, Miquel and Wunsch}]{le11}
\textsc{Lechner, M.}, \textsc{Miquel, R.} and \textsc{Wunsch, C.} (2011).
  {Long-Run Effects of Public Sector Sponsored Training in West Germany}.
  \textit{Journal of the European Economic Association}, \textbf{9}~(4),
  742–784.

\bibitem[{Lechner and Strittmatter(2019)}]{le19}
\textsc{---} and \textsc{Strittmatter, A.} (2019). {Practical Procedures to
  Deal with Common Support Problems in Matching Estimation}.
  \textit{Econometric Reviews}, \textbf{38}~(2), 193--207.

\bibitem[{{New York Times}(2015)}]{ne15}
\textsc{{New York Times}} (2015). {Volkswagen Says 11 Million Cars Worldwide
  Are Affected in Diesel Deception}. p. 23.09.2015.

\bibitem[{Peterson and Schneider(2014)}]{pe14}
\textsc{Peterson, J.} and \textsc{Schneider, H.} (2014). {Adverse Selection in
  the Used-Car Market: Evidence from Purchase and Repair Patterns in the
  Consumer Expenditure Survey}. \textit{RAND Journal of Economics},
  \textbf{45}~(1), 140--154.

\bibitem[{Peterson and Schneider(2016)}]{pe16}
\textsc{---} and \textsc{---} (2016). {Beautiful Lemons: Adverse Selection in
  Durable-Goods Markets with Sorting}. \textit{Management Science},
  \textbf{63}~(9), 3111–3127.

\bibitem[{Rosenbaum and Rubin(1983)}]{ro83}
\textsc{Rosenbaum, P.} and \textsc{Rubin, D.} (1983). {The Central Role of the
  Propensity Score in Observational Studies for Causal Effects}.
  \textit{Biometrika}, \textbf{70}~(1), 41–55.

\bibitem[{Rubin(1979)}]{ru79}
\textsc{Rubin, D.} (1979). {Using Multivariate Matched Sampling and Regression
  Adjustment to Control Bias in Observational Studies}. \textit{Journal of the
  American Statistical Associatio}, \textbf{74}~(2), 318–328.

\bibitem[{Sallee \textit{et~al.}(2016)Sallee, West and Fan}]{sa16}
\textsc{Sallee, J.}, \textsc{West, S.} and \textsc{Fan, W.} (2016). {Do
  Consumers Recognize the Value of the Fuel Economy? Evidence from Used Car
  Prices and Gasoline Price Fluctuations}. \textit{Journal of Public
  Economics}, \textbf{135}, 61--73.

\bibitem[{Spiegel(2016)}]{sp16}
\textsc{Spiegel} (2016). {Bosch hat Schummelsoftware nicht nur an VW
  geliefert}. pp. Issue 17, 23.04.2016.

\bibitem[{Stanwick and Stanwick(2017)}]{st17}
\textsc{Stanwick, P.} and \textsc{Stanwick, S.} (2017). {Volkswagen Emissions
  Scandal: The Perils of Installing Illegal Software}. \textit{International
  Review of Management and Business Research}, \textbf{6}~(1), 18--24.

\bibitem[{Tadelis and Zettelmeyer(2015)}]{ta15}
\textsc{Tadelis, S.} and \textsc{Zettelmeyer, F.} (2015). {Information
  Disclosure as a Matching Mechanism: Theory and Evidence from a Field
  Experiment}. \textit{American Economic Review}, \textbf{105}~(2), 886--905.

\bibitem[{Thompson \textit{et~al.}(2014)Thompson, Carder, Besch, Thiruvengadam
  and Kappanna}]{th14}
\textsc{Thompson, G.}, \textsc{Carder, D.}, \textsc{Besch, M.},
  \textsc{Thiruvengadam, A.} and \textsc{Kappanna, H.} (2014). {In-Use
  Emissions Testing of Light-Duty Diesel Vehicles in the United States}.
  \textit{Final Report prepared for the International Council on Clean
  Transportation (ICCT)}.

\bibitem[{Yang \textit{et~al.}(2015)Yang, Franco, Campestrini, German and
  Mock}]{ya15}
\textsc{Yang, L.}, \textsc{Franco, V.}, \textsc{Campestrini, A.},
  \textsc{German, J.} and \textsc{Mock, P.} (2015). {$NO_{X}$ Control
  Technologies for Euro 6 Diesel Passenger Cars: Market Penetration and
  Experimental Performance Assessment}. \textit{Whitepaper, National Academies
  of Sciences, Engineering, and Medicine}.

\end{thebibliography}

\clearpage
\section*{Figures and Tables}

 \clearpage
  
\begin{table}[h!]  
\caption{Trends in the German car market.} \label{tab1}\centering
\begin{tabular}{lcccccc}\hline\hline
    Year  & \multicolumn{2}{c}{Registrations of New Cars} & \multicolumn{2}{c}{Ownership Transfers} & \multicolumn{2}{c}{Decommissions} \\
          & Gasoline & Diesel & Gasoline & Diesel & Gasoline & Diesel \\
          & (in 100,000) & (in 100,000) & (in 100,000) & (in 100,000) & (in 100,000) & (in 100,000) \\ \hline
    2008  & 17.0  & 13.6  & 46.8  & 13.8  & 53.1  & 21.8 \\
    2009  & 26.1  & 11.7  & 43.0  & 16.6  & 60.1  & 20.1 \\
    2010  & 16.7  & 12.2  & 46.9  & 16.7  & 50.5  & 20.7 \\
    2011  & 16.5  & 15.0  & 49.5  & 17.7  & 54.2  & 23.1 \\
    2012  & 15.6  & 14.9  & 48.1  & 19.8  & 54.7  & 24.5 \\
    2013  & 15.0  & 14.0  & 48.5  & 21.4  & 54.9  & 25.4 \\
    2014  & 15.3  & 14.5  & 47.3  & 22.2  & 53.3  & 26.8 \\
    2015  & 16.1  & 15.4  & 48.4  & 23.8  & 53.7  & 28.5 \\
    2016  & 17.5  & 15.4  & 48.3  & 24.6  & 53.5  & 30.4 \\
    2017  & 19.9  & 13.4  & 47.8  & 23.9  & 52.7  & 32.1 \\\hline\hline
    \end{tabular}
\parbox{\textwidth}{\footnotesize \emph{Notes:} This table reports the number of new car registrations, ownership transfers, and decommissions in Germany by fuel type and year. Owners decommission their cars because they wish to sell or scrap them at a later point in time. The peak of new car registrations in 2009 is from the ‘cash for clunkers’ program. Data source: \cite{kr17}.}
\end{table}

  
\begin{table}[h!]  
\caption{Vehicle classes.} \label{tab2}\centering
    \begin{tabular}{|p{0.17\textwidth}|p{0.65\textwidth}|p{0.16\textwidth}|} \hline
      Vehicle class & Considered car models & Observations \\\hline
    Small & VW Polo, Opel Corsa, Ford Fiesta, Renault Clio, Peugeot 208, Fiat Punto, Toyota Yaris & 151,346 (14\%) \\\hline
     Compact & VW Golf, Mercedes A-class, BMW series 1, Opel Astra, Ford Focus, Renault Megane, Peugeot 308, Toyota Auris & 417,523 (38\%) \\\hline
     Medium-sized & VW Passat, Mercedes C-class, BMW series 3, Opel Insignia, Ford Mondea, Peugeot 508, Toyota Avensis & 286,205 (26\%) \\\hline
     Minivan & VW Sharan, VW Touran, Mercedes B-class, BMW series 2, Opel Zafira, Ford Galaxy, Ford C-Max, Renault Scenic, Peugeot 5008, Fiat Doblo, Toyota Verso & 148,612 (13\%) \\\hline
    SUV   & VW Tiguan, Mercedes GLA, Mercedes GLK, BMW X1, Opel Mokka, Ford Kuga, Renault Kadjar, Toyota RAV4 & 104,880  (9\%) \\\hline
    \end{tabular}
\end{table}


\begin{table}[h!]    \vspace{-0.4cm}
\caption{Descriptive statistics.} \label{tab3}\centering
\begin{tabularx}{\textwidth}{Xccccc}\hline\hline
    Variables & \multicolumn{2}{c}{VW vehicles} & \multicolumn{2}{c}{Vehicles from } & Standardized \\
         & \multicolumn{2}{c}{} & \multicolumn{2}{c}{other carmakers} & Difference \\
          & Mean  & Std. Err. & Mean  & Std. Err. &  \\
    \cline{2-6}      & (1)   & (2)   & (3)   & (4)   & (5) \\ \hline
    Diesel & 0.54  & -     & 0.52  & -     & 2.98 \\
    First asking price (in thsd. euros) & 17.1  & 8.0   & 16.5  & 9.8   & 0.06 \\\hline
    Car age (in years) & 3.08  & 3.09  & 3.76  & 3.31  & 21.13 \\
    Mileage (in 1,000 km) & 61.3  & 52.4  & 66.6  & 51.5  & 10.2 \\
    Full service history & 0.76  & -     & 0.61  & -     & 32.03 \\
    Warranty & 0.08  & -     & 0.11  & -     & 10.42 \\
    New general inspection & 0.18  & -     & 0.23  & -     & 12.13 \\
    Duration until next general  inspection above 1 year & 0.57  & -     & 0.49  & -     & 14.35 \\
    Private car seller & 0.10  & -     & 0.13  & -     & 7.32 \\\hline
    \multicolumn{6}{c}{Vehicle class} \\\hline
Small & 0.12 & - & 0.15  & -     & 6.79 \\
  Compact & 0.49 & - & 0.31  & -     & 36.70 \\
Middle-sized & 0.19 & - & 0.30  & -     & 27.85 \\
Minivan & 0.12 & - & 0.14  & -     & 4.49 \\
SUV & 0.08 & - & 0.10  & -     & 6.83 \\\hline
 \multicolumn{6}{c}{Car body}\\\hline
    Limousine & 0.42  & -     & 0.40  & -     & 3.23 \\
    Estate & 0.26  & -     & 0.29  & -     & 6.95 \\
    Minivan & 0.10  & -     & 0.08  & -     & 6.16 \\
    SUV   & 0.08  & -     & 0.09  & -     & 4.48 \\
    Small & 0.06  & -     & 0.07  & -     & 3.54 \\
    Sport & 0.03  & -     & 0.05  & -     & 10.73 \\
    Others & 0.05  & -     & 0.01  & -     & 22.09 \\\hline
    \multicolumn{6}{c}{Environmental standards} \\\hline
    Below Euro 4 & 0.07  & -     & 0.11  & -     & 15.90 \\
    Euro 4 & 0.13  & -     & 0.22  & -     & 24.29 \\
    Euro 5 & 0.53  & -     & 0.49  & -     & 8.69 \\
    Euro 6 & 0.27  & -     & 0.18  & -     & 22.44 \\
    Green particulate matter tag & 0.87  & -     & 0.82  & -     & 14.83 \\\hline
    \multicolumn{6}{c}{Fuel consumption (combined)} \\\hline
    Below 5 liters/100km & 0.35  & -     & 0.29  & -     & 12.89 \\
    Between 5 and 6 liters/100km & 0.30  & -     & 0.30  & -     & 0.74 \\
    Above 6 liters/100km & 0.27  & -     & 0.30  & -     & 7.41 \\
    Not reported & 0.09  & -     & 0.11  & -     & 7.79 \\\hline
    Observations & \multicolumn{2}{c}{427,286} & \multicolumn{2}{c}{681,280} &  \\\hline\hline
    \end{tabularx}
\parbox{\textwidth}{\footnotesize \emph{Notes:} This table reports the mean car characteristics of vehicles from VW and other car makes. The other car makes include Mercedes, BMW, Opel, Ford, Renault, Peugeot, Fiat, and Toyota cars. Columns (2) and (4) document standard errors of continuous variables. Column (5) reports the standardized difference. The standardized difference of variable $X$ between samples $A$ and $B$ is defined as $SD = 100 \cdot |\bar{X}_A-\bar{X}_B|/\sqrt{1/2\left(Var(X_A) + Var(X_B)\right)}$,
where $\bar{X}_A$ denotes the mean of sample $A$ and $\bar{X}_B$ denotes the mean of sample $B$. \cite{ro83} consider an absolute standardized difference of more than 20 as ‘large’.}
\end{table}

  
\begin{table}[h!]  
\caption{Matched difference in the share of used diesel vehicles by emission standard.} \label{tab4}\centering
\begin{tabularx}{\textwidth}{Xcccccccc}\hline\hline
            & \multicolumn{7}{c}{Month after disclosure of the emission scandal} \\
          & 1.    & 2.    & 3.    & 4.    & 5.    & 6.    & 7. \\
      \cline{1-8}    & (1)   & (2)   & (3)   & (4)   & (5)   & (6)   & (7) \\\hline
  Below Euro 5  & -0.014 & -0.018 & -0.004 & 0.006 & -0.015 & -0.005 & 0.005 \\
  & (0.014) & (0.015) & (0.015) & (0.016) & (0.015) & (0.015) & (0.016) \\
  Euro 5 & 0.0001 & 0.019** & 0.01  & 0.018** & 0.038*** & 0.024*** & 0.004 \\
          & (0.007) & (0.007) & (0.007) & (0.008) & (0.007) & (0.007) & (0.008) \\
Euro 6 & -0.036 & -0.012 & 0.01  & 0.025 & -0.03 & -0.061*** & -0.032 \\
          & (0.024) & (0.022) & (0.021) & (0.023) & (0.022) & (0.02) & (0.021) \\ \hline \hline
     \end{tabularx}
\parbox{\textwidth}{\footnotesize \emph{Notes:}
This table reports the matched difference-in-differences of the used car diesel share for the first seven months after the disclosure of the VW emission scandal by emission standard. Defeat devices were only disclosed for VW diesel vehicles regulated under the Euro 5 emission standard. Diesel and gasoline cars are matched at the individual car level such that they share the same characteristics. The first matched difference is between the time before and after the disclosure of the VW emission scandal. The second matched difference is between VW and other car makes. The other car makes include Mercedes, BMW, Opel, Ford, Renault, Peugeot, Fiat, and Toyota cars. The standard errors in parentheses are calculated with a non-parametric bootstrap (499 replications). The asterisks indicate 1\% (***), 5\% (**), and 10\% (*) significance levels. }
\end{table}

  
\begin{table}[h!]  
\caption{Matched difference in the first asking price of diesel vehicles by emission standard.} \label{tab5}\centering
\begin{tabularx}{\textwidth}{Xcccccccc}\hline\hline
            & \multicolumn{7}{c}{Month after disclosure of the emission scandal} \\
          & 1.    & 2.    & 3.    & 4.    & 5.    & 6.    & 7. \\
      \cline{2-8}    & (1)  )& (2)   & (3)   & (4)   & (5)   & (6)   & (7) \\\hline
 Below Euro 5 & 148   & -147  & -106  & 1,242*** & 67    & -648** & -269 \\
       & (268)   & (272)   & (261)   & (280)   & (246)   & (238)   & (248) \\
 Euro 5 & -17   & -287** & -611*** & -570*** & -521*** & -258*  & -325* \\
      &   (133)   & (141)   & (144)   & (152)   & (149)   & (148)   & (155) \\
    Euro 6 & 1,634*** & 998** & 1,299*** & 2,653*** & 792*  & 478   & 640 \\
   &  (426)   & (461)   & (436)   & (508)   & (448)   & (412)   & (441) \\\hline \hline
     \end{tabularx}
\parbox{\textwidth}{\footnotesize \emph{Notes:} 
 This table reports the matched difference-in-differences in the first asking price (in euros) of used diesel vehicles for the first seven months after the disclosure of the VW emission scandal by emission standard. Defeat devices were only disclosed for VW diesel vehicles regulated under the Euro 5 emission standard. Diesel cars are matched at the individual car level such that they share the same characteristics. The first matched difference is between the time before and after the disclosure of the VW emission scandal. The second matched difference is between VW and other car makes. The other car makes include Mercedes, BMW, Opel, Ford, Renault, Peugeot, Fiat, and Toyota diesel cars. The standard errors in parentheses are calculated with a non-parametric bootstrap (499 replications). The asterisks indicate 1\% (***), 5\% (**), and 10\% (*) significance levels. }
\end{table}

  
\begin{table}[h!]  
\caption{Matched difference in the share of used diesel vehicles by vehicle class.} \label{tab6}\centering
\begin{tabularx}{\textwidth}{Xcccccccc}\hline\hline
            & \multicolumn{7}{c}{Month after disclosure of the emission scandal} \\
          & 1.    & 2.    & 3.    & 4.    & 5.    & 6.    & 7. \\
      \cline{2-8}    & (1)   & (2)   & (3)   & (4)   & (5)   & (6)   & (7) \\\hline
    Small & -0.03* & -0.002 & -0.023 & -0.007 & 0.005 & 0.015 & -0.012 \\
          & (0.015) & (0.016) & (0.016) & (0.016) & (0.016) & (0.015) & (0.016) \\
    Compact & -0.01 & 0.002 & 0.002 & 0.0002 & -0.003 & -0.009 & 0.023* \\
          & (0.013) & (0.013) & (0.013) & (0.014) & (0.013) & (0.014) & (0.013) \\
    Medium-sized & -0.001 & 0.002 & 0.007 & 0.018 & 0.029** & 0.008 & 0.015 \\
          & (0.011) & (0.011) & (0.011) & (0.012) & (0.011) & (0.011) & (0.013) \\
    Minivan & -0.009 & 0.001 & 0.026 & 0.016 & 0.053*** & 0.068*** & 0.035 \\
          & (0.019) & (0.019) & (0.02) & (0.02) & (0.02) & (0.022) & (0.023) \\
    SUV & 0.004 & 0.001 & 0.004 & 0.053** & 0.071*** & 0.052** & 0.101*** \\
          & (0.021) & (0.021) & (0.022) & (0.021) & (0.024) & (0.023) & (0.024) \\ \hline \hline
     \end{tabularx}
\parbox{\textwidth}{\footnotesize \emph{Notes:} This table reports the matched difference-in-differences of the used car diesel share for the first seven months after the disclosure of the VW emission scandal by vehicle class. Diesel and gasoline cars are matched at the individual car level such that they share the same characteristics. The first matched difference is between the time before and after the disclosure of the VW emission scandal. The second matched difference is between VW and other car makes. The other car makes include Mercedes, BMW, Opel, Ford, Renault, Peugeot, Fiat, and Toyota cars. The standard errors in parentheses are calculated with a non-parametric bootstrap (499 replications). The asterisks indicate 1\% (***), 5\% (**), and 10\% (*) significance levels. }
\end{table}

  
\begin{table}[h!]  
\caption{Matched difference in the first asking price of diesel vehicles by vehicle class.} \label{tab7}\centering
\begin{tabularx}{\textwidth}{Xcccccccc}\hline\hline
            & \multicolumn{7}{c}{Month after disclosure of the emission scandal} \\
          & 1.    & 2.    & 3.    & 4.    & 5.    & 6.    & 7. \\
      \cline{2-8}    & (1)   & (2)   & (3)   & (4)   & (5)   & (6)   & (7) \\\hline
     Small & 85    & 194   & -227  & 149   & 122   & -154  & -287 \\
          & (293)   & (291)   & (287)   & (279)   & (303)   & (283)   & (273)  \\
    Compact & -54   & -414** & 298   & -406* & -505** & -524** & -455** \\
          & (207)   & (178)   & (223)   & (220)   & (193)   & (221)   & (204) \\
 Medium-sized & 379*   & 165   & -67   & 38    & -59   & 439* & 233 \\
          & (231)   & (223)   & (240)   & (240)   & (227)   & (227)   & (232)  \\
    Minivan & 221   & -75   & -46   & 531*  & 487   & 306   & 373 \\
          &     (302)   & (290)   & (306)   & (304)   & (316)   & (297)   & (305)  \\
   SUV & -322  & -850** & -421  & -244  & -451  & -502  & -316 \\
          &   (382)   & (383) & (387)   & (398)   & (380)   & (389)   & (390)  \\ \hline \hline
     \end{tabularx}
\parbox{\textwidth}{\footnotesize \emph{Notes:} 
 This table reports the matched difference-in-differences in the first asking price (in euros) of used diesel vehicles for the first seven months after the disclosure of the VW emission scandal by vehicle class. Diesel cars are matched at the individual car level such that they share the same characteristics. The first matched difference is between the time before and after the disclosure of the VW emission scandal. The second matched difference is between VW and other car makes. The other car makes include Mercedes, BMW, Opel, Ford, Renault, Peugeot, Fiat, and Toyota diesel cars. The standard errors in parentheses are calculated with a non-parametric bootstrap (499 replications). The asterisks indicate 1\% (***), 5\% (**), and 10\% (*) significance levels. }
\end{table}

  
\begin{table}[h!]  
\caption{Matched difference in the share of used diesel vehicles by seller type.} \label{tab8}\centering
\begin{tabularx}{\textwidth}{Xcccccccc}\hline\hline
            & \multicolumn{7}{c}{Month after disclosure of the emission scandal} \\
          & 1.    & 2.    & 3.    & 4.    & 5.    & 6.    & 7. \\
      \cline{2-8}    & (1)   & (2)   & (3)   & (4)   & (5)   & (6)   & (7) \\\hline
    Private seller & -0.011 & -0.03* & -0.043** & -0.033* & -0.022 & -0.03* & -0.022 \\
          & (0.018) & (0.017) & (0.019) & (0.018) & (0.018) & (0.017) & (0.018) \\
    Professional car  & -0.017** & -0.0004 & 0.014** & 0.035*** & 0.019*** & 0.008 & 0.014** \\
      dealer    & (0.007) & (0.007) & (0.007) & (0.007) & (0.007) & (0.007) & (0.007) \\
   Professional car  & -0.027 & -0.016 & 0.005 & 0.013 & 0.064*** & 0.01  & -0.011 \\
   dealer providing warranty       & (0.024) & (0.023) & (0.025) & (0.025) & (0.025) & (0.025) & (0.024) \\
    Professional car & -0.011 & 0.001 & 0.015* & 0.032*** & 0.028*** & 0.012 & 0.016** \\
    dealer not providing warranty       & (0.008) & (0.007) & (0.008) & (0.007) & (0.007) & (0.007) & (0.007) \\ \hline \hline
     \end{tabularx}
\parbox{\textwidth}{\footnotesize \emph{Notes:} This table reports the matched difference-in-differences of the used car diesel share for the first seven months after the disclosure of the VW emission scandal by seller type. Diesel and gasoline cars are matched at the individual car level such that they share the same characteristics. The first matched difference is between the time before and after the disclosure of the VW emission scandal. The second matched difference is between VW and other car makes. The other car makes include Mercedes, BMW, Opel, Ford, Renault, Peugeot, Fiat, and Toyota cars. The standard errors in parentheses are calculated with a non-parametric bootstrap (499 replications). The asterisks indicate 1\% (***), 5\% (**), and 10\% (*) significance levels.}
\end{table}

  
\begin{table}[h!]  
\caption{Matched difference in the first asking price of diesel vehicles by seller type.} \label{tab9}\centering
\begin{tabularx}{\textwidth}{Xcccccccc}\hline\hline
            & \multicolumn{7}{c}{Month after disclosure of the emission scandal} \\
          & 1.    & 2.    & 3.    & 4.    & 5.    & 6.    & 7. \\
      \cline{2-8}    & (1)   & (2)   & (3)   & (4)   & (5)   & (6)   & (7) \\\hline
       Private seller & 346   & 475   & 771** & 481   & 398   & 536   & 393 \\
          & (324)   & (324)   & (334)   & (347)   & (279)   & (338)   & (314)  \\
    Professional car  dealer & 179   & -254* & -32   & -3    & -310** & -57   & 105 \\
    & (128)   & (130)   & (127)   & (147)   & (143)   & (148)   & (143) \\
    Professional car dealer providing & 375   & -22   & -640* & 430   & -206  & 22    & 488 \\
     warranty & (381)   & (394)   & (359)   & (377)   & (354)   & (359)   & (371)  \\
    Professional car dealer not   & -6    & -240  & 35    & -255  & -194  & -286** & 50 \\
     providing warranty &  (156)   & (152)   & (158)   & (188)   & (155)   & (140)   & (154) \\ \hline \hline
     \end{tabularx}
\parbox{\textwidth}{\footnotesize \emph{Notes:} 
 This table reports the matched difference-in-differences in the first asking price (in euros) of used diesel vehicles for the first seven months after the disclosure of the VW emission scandal by seller type. Diesel cars are matched at the individual car level such that they share the same characteristics. The first matched difference is between the time before and after the disclosure of the VW emission scandal. The second matched difference is between VW and other car makes. The other car makes include Mercedes, BMW, Opel, Ford, Renault, Peugeot, Fiat, and Toyota diesel cars. The standard errors in parentheses are calculated with a non-parametric bootstrap (499 replications). The asterisks indicate 1\% (***), 5\% (**), and 10\% (*) significance levels.}
\end{table}

\end{document}